\documentclass[lettersize,journal]{IEEEtran}
\usepackage{amsmath,amssymb,amsfonts}
\usepackage{algorithmic}
\usepackage{algorithm}
\usepackage{array}
\usepackage{textcomp}
\usepackage{stfloats}
\usepackage{url}
\usepackage{verbatim}
\usepackage{graphicx}
\usepackage{cite}
\usepackage[breaklinks=true,colorlinks,bookmarks=false]{hyperref}
\usepackage{color}
\usepackage{colortbl}
\usepackage{makecell}
\usepackage{threeparttable}
\usepackage{overpic}
\usepackage{multirow}
\usepackage{rotating}
\usepackage{xspace}

\usepackage{booktabs}
\usepackage{adjustbox}

\def\ie{\emph{i.e.}}
\def\eg{\emph{e.g.}}

\def\etal{\emph{et al.}}

\newcommand{\figref}[1]{Fig. \ref{#1}}
\newcommand{\tabref}[1]{Tab. \ref{#1}}

\newcommand{\secref}[1]{\S\ref{#1}}

\def\RevDataNum{63}
\def\RevPaperNum{137}
\def\ourmodel{ColonGPT}
\def\ourdata{ColonINST}

\definecolor{mygray4}{gray}{0.55}
\definecolor{mygray5}{gray}{0.9}
\newcommand{\tgray}[1]{\textcolor{mygray4}{#1}}
\newcommand{\pink}[1]{\textcolor{magenta}{#1}}
\newcommand{\revA}[1]{\textcolor{black}{#1}}

\begin{document}

\title{Frontiers in Intelligent Colonoscopy}

\author{Ge-Peng Ji,~Jingyi Liu,~Peng Xu,~Nick Barnes,~Fahad Shahbaz Khan,~Salman Khan,~Deng-Ping Fan
\thanks{*Corresponding author: Deng-Ping Fan (\href{mailto:dengpfan@gmail.com}{dengpfan@gmail.com}).}
\thanks{D. Fan is with the Nankai Institute of Advanced Research (SHENZHEN FUTIAN), Guangdong, China, and also with the College of Computer Science \& VCIP, Nankai University, Tianjin, China.}
\thanks{G. Ji and N. Barnes are with the School of Computing, Australian National University, Canberra, Australia.}
\thanks{J. Liu is with the Graduate School of Science and Technology, Keio University, Yokohama, Japan.}
\thanks{P. Xu is with the Department of Electronic Engineering, Tsinghua University, Beijing, China.}
\thanks{F. Khan and S. Khan are with Mohamed bin Zayed University of Artificial Intelligence, Abu Dhabi, UAE.}
}



\markboth{Journal of \LaTeX\ Class Files,~Vol.~14, No.~8, August~2021}%
{Shell \MakeLowercase{\textit{et al.}}: A Sample Article Using IEEEtran.cls for IEEE Journals}


\maketitle

\begin{abstract}
Colonoscopy is currently one of the most sensitive screening methods for colorectal cancer.
This study investigates the frontiers of intelligent colonoscopy techniques and their prospective implications for multimodal medical applications.
With this goal, we begin by assessing the current data-centric and model-centric landscapes through four tasks for colonoscopic scene perception, including classification, detection, segmentation, and vision-language understanding. 
Our assessment reveals domain-specific challenges and underscores the need for further multimodal research in colonoscopy. To address these gaps, we establish three foundational initiatives: a large-scale multimodal instruction tuning dataset \ourdata, a colonoscopy-designed multimodal language model~\ourmodel,
and a multimodal benchmark. To facilitate continuous advancements in this rapidly evolving field, we provide a public website for the latest updates: \url{https://github.com/ai4colonoscopy/IntelliScope}.
\end{abstract}

\begin{IEEEkeywords}
Colonoscopy, Survey, Vision-language, Multimodal language model, Multimodal benchmark, Abdomen, Healthcare AI.
\end{IEEEkeywords}

\section{Introduction}
\label{sec:introduction}
\IEEEPARstart{D}{espite} declining colorectal cancer (CRC) rates in high-income countries, it remains the third most diagnosed cancer worldwide and is increasing in developing countries \cite{ENG2024}. As an efficient method for CRC screening, colonoscopy utilises a flexible camera-equipped tube to visually examine the colon's interior. As illustrated in Fig. \hyperlink{fig1_a}{(1-a)}, this clinical procedure also facilitates intervention with specialised instruments such as snares, forceps, and cautery devices to remove precancerous growths, such as serrated and adenomatous polyps. A recent study \cite{wallace2022impact} indicates that incorporating artificial intelligence (AI) into colonoscopy reduces the miss rate of colorectal neoplasia by approximately 50\% compared to traditional methods. This success motivates us to investigate the \textit{frontiers in intelligent colonoscopy}. 

Colonoscopy, an endoscopic optical imaging technique, usually presents challenges such as non-uniform illumination and homogeneity of visual patterns, that differ from those of general-purpose imaging data, \eg, ImageNet \cite{deng2009imagenet}, due to the complex and folded anatomy of the colon.
This suggests that special methods are needed to interpret the colonoscopic data. In response, we begin with an investigation of the latest intelligent techniques for colonoscopy, assessing the current landscape to sort out domain-unique challenges and underexplored areas. Our analysis reveals that multimodal research in colonoscopy remains largely untapped. To bridge this gap, we contribute the following three efforts to the community, as illustrated in Fig. \hyperlink{fig1_b}{(1-b)}.

\begin{figure}[t!]
\centering
\includegraphics[width=\linewidth]{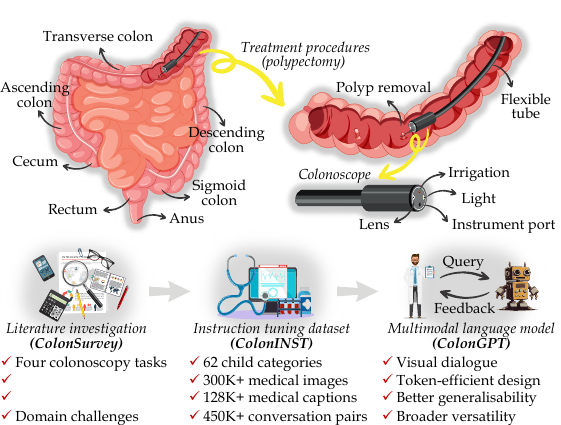}
\put(-245.7,17.9){\fontsize{7pt}{12pt}\selectfont \RevDataNum~colonoscopy datasets}
\put(-245.8,10.2){\fontsize{7pt}{12pt}\selectfont \RevPaperNum~related models}
\put(-253,185.5){\scriptsize \hypertarget{fig1_a}{\textbf{(a) Illustration of a colonoscope inside large intestine (colon)}}}
\put(-253,82.5){\scriptsize \hypertarget{fig1_b}{\textbf{(b) Highlights of this study}}}
\vspace{-8pt}
\caption{\textbf{Overview of colonoscopy and study highlights:} We depict (a) the anatomy of the large intestine (colon) within the digestive tract, the polypectomy procedure during colonoscopy examination, and the colonoscope components. 
(b) summarises three highlights of this study.
}
\label{fig:teaser_figure}
\end{figure}

\begin{figure*}[t!]
\centering
\includegraphics[width=\linewidth]{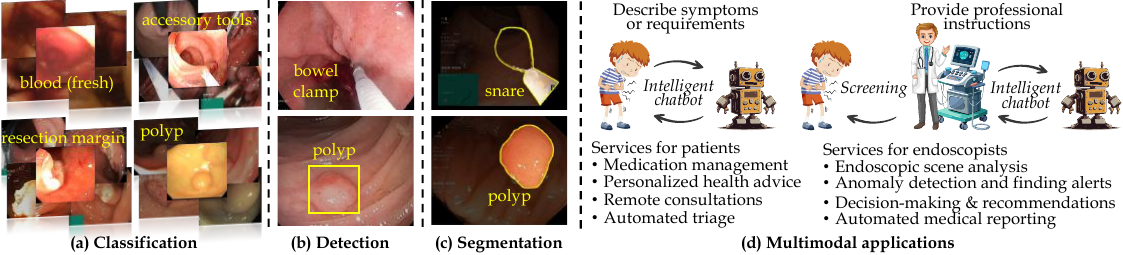}
\vspace{-18pt}
\caption{\textbf{Colonscopic scene perception from visual to multimodal perspectives.} 
In clinical practice, purely visual tasks, including (a) classification, (b) detection, and (c) segmentation, are applied to identify 
targets of interest such as polyps and instruments. 
%
(d) Multimodal applications
improve colonoscopy procedures by performing interactive, user-driven tasks aligned with clinical needs. The chatbot provides personalised advice, automated reporting, and streamline procedural workflows.
}
\label{fig:task_illustration}
\end{figure*}

\noindent\textbf{Contribution.} 
\textit{(a)} We investigate the latest research progress in four colonoscopic scene perception tasks (refer to \figref{fig:task_illustration}) from both data-centric and model-centric perspectives. Our investigation summarises key features of \RevDataNum\! datasets and \RevPaperNum\! representative deep learning techniques published since 2015. Additionally, we highlight emerging trends and opportunities for future study.
\textit{(b)} We introduce \ourdata, a pioneering instruction tuning dataset tailored for multimodal research, aimed at instructing models to execute user-driven tasks interactively. Assembled from 19 publicly available sources, the \ourdata\!
dataset contains 303,001 colonoscopy images across 62 sub-categories, reflecting diverse scenarios encountered in colonoscopy procedures. We expand these visual samples in two aspects. First, we leverage the multimodal AI chatbot, GPT-4V \cite{openai2023chatgpt4v}, to generate 128,620 medical captions. Second, we restructure 450,724 human-machine conversations for multimodal adaptation.
\textit{(c)} Leveraging the instruction tuning data, we build a multimodal language model, \ourmodel\, to assist endoscopists through interactive dialogues. To ensure reproducibility for average community users, we implement ColonGPT in a resource-friendly way, using a 0.4B-parameter visual encoder SigLIP-SO \cite{zhai2023sigmoid} and a 1.3B-parameter lightweight language model Phi1.5 \cite{li2023textbooks}. Unlike previous vision-language (VL) linking methods \cite{chen2023minigptv2,liu2024llavav15,he2024bunny} that employ multilayer perceptrons to handle all tokens equally from the visual encoder, we propose a multigranularity adapter to selectively sample visual tokens according to their significance. This strategy reduces the visual tokens to $\sim$34\% of the original number without compromising performance, securing the top spot in our newly created multimodal benchmark across three tasks. Importantly, our \ourmodel\! can be trained within \revA{seven hours on two NVIDIA H200 GPUs}, facilitating rapid proof-of-concept development for subsequent research.
 

%
\noindent\textbf{Scope.} 
This study differs from previous works in several aspects. Earlier surveys on traditional \cite{prasath2016polyp} and deep learning \cite{taha2017automatic,sanchez2020deep,pacal2020comprehensive,munzer2018content} methods conducted before 2020 are now out of date.
Although a recent study \cite{taghiakbari2021artificial} explores various applications of colonoscopy, such as quality analysis and anomaly detection, it lacks numerical validation.
Other benchmarks \cite{psnbi2k,wu2024colorectal,mei2023survey} are limited to specific narrow research subfields.
By contrast, we delve into four tasks related to colonoscopic scene perception and evaluate their current state to identify key challenges and under-researched areas. Importantly, our vision goes beyond by laying the foundations for the upcoming advancements in the multimodal domain. To this end, we undertake three initiatives to community: a multimodal instruction tuning dataset, a multimodal language model, and a multimodal benchmark.

\noindent\textbf{Organisation.} The remaining sections are structured as follows: \secref{sec:background} provides a historical background and discusses the unique challenges of the domain. In \secref{sec:colonoscopy_related_datasets}, we investigate {\RevDataNum} datasets related to colonoscopy, followed by a survey of {\RevPaperNum} deep models in \secref{sec:colonoscopy_models}. In \secref{sec:multimodal_solution}, we introduce three initiatives towards the multimodal era: the creation of \ourdata~dataset, the technical details of \ourmodel~model, and a comparative multimodal benchmark along with ablative analyses. Finally, this paper is concluded in \secref{sec:conclusion}.

\begin{table*}[t!]
\centering
\fontsize{6.2pt}{7pt}\selectfont
\renewcommand{\arraystretch}{0.9}
\renewcommand{\tabcolsep}{0.6mm}
\caption{\textbf{Data statistics for colonoscopy datasets.}
The columns include: 
number of images (\#IMG) and videos (\#VID), classification tag (Cls), bounding box (Bbx), segmentation mask (Seg), text (Tx). 
The categories not related to colonoscopy, such as stomach and esophagitis, are marked in \tgray{grey}.}
\label{tab:colonoscopy_datasets}
\begin{tabular}{r|r|r|r|cccc|l|c}
    Dataset & Publication & \#IMG & \#VID & Cls & Bbx & Seg & Tx & Number of categories (\#C) $\rightarrow$ Category names & URL \\
\hline
    CVC-ColonDB \cite{bernal2012towards} & PR'12 
    & 300 & 15 
    &  &  & \checkmark &
    & \#C1 $\rightarrow$ polyp 
    & \href{http://vi.cvc.uab.es/colon-qa/cvccolondb/}{Link} \\
    ETIS-Larib~\cite{silva2014toward} & CARS'14 
    & 196 & - 
    & & & \checkmark &
    & \#C2 $\rightarrow$ polyp, non-polyp
    & \href{https://polyp.grand-challenge.org/ETISLarib/}{Link} \\ 
%
    CVC-ClinicDB \cite{bernal2015wm} & CMIG'15 
    & 612 & 31
    & & & \checkmark &
    & \#C1 $\rightarrow$ polyp 
    & \href{https://polyp.grand-challenge.org/CVCClinicDB/}{Link} \\
    ASU-Mayo \cite{tajbakhsh2015automated} & TMI'15 
    & 36,458 & 38 
    & \checkmark & & \checkmark &
    & \#C2 $\rightarrow$ polyp, non-polyp 
    & \href{https://polyp.grand-challenge.org/AsuMayo/}{Link} \\
    Ye \etal \cite{ye2016online} & MedIA'16 
    & 7,894 & 10 
    & \checkmark & \checkmark &  &
    & \#C2 $\rightarrow$ polyp, non-instance 
    & \href{http://hamlyn.doc.ic.ac.uk/vision/}{Link} \\
    Deeba \etal \cite{deeba2016automated} & IJCNN'16 
    & 100 & -
    & \checkmark &  &  &
    & \#C2 $\rightarrow$ bleeding, non-bleeding
    & - \\
    CU-ColonDB \cite{zhang2016automatic} & JBHI'16 
    & 1,930 & - 
    & \checkmark & & &
    & \#C3 $\rightarrow$ hyperplasia polyps, adenomatous polyps, non-polyp 
    & - \\
    ColonoscopicDS \cite{mesejo2016computer} & TMI'16 
    & - & 76 
    & \checkmark & & &
    & \#C3 $\rightarrow$ serrated adenomas, hyperplastic lesions, adenoma
    & \href{http://www.depeca.uah.es/colonoscopy_dataset/}{Link}\\ 
    CVC-ClinicVideoDB \cite{angermann2017towards} & MICCAIw'17 
    & 10,924 & 18 
    & \checkmark & \checkmark & \checkmark &
    & \#C2 $\rightarrow$ polyp, non-polyp 
    & \href{https://endovissub2017-giana.grand-challenge.org/}{Link} \\ 
    Kvasir \cite{pogorelov2017kvasir} 
    & MMSys'17
    & 8,000 
    & -
    & \checkmark  & & &
    & \#C8 $\rightarrow$ cecum, polyps, ulcerative colitis, dyed and lifted polyp, dyed resection margins, \tgray{Z-line, pylorus, esophagitis}
    & \href{https://datasets.simula.no/kvasir/}{Link} \\
    Nerthus \cite{pogorelov2017nerthus} & MMSys'17 
    & 5,525 & 21 
    & \checkmark & & &
    & \#C4 $\rightarrow$ BBPS (Boston-Bowel-Preparation-Scale) 0/1/2/3 
    & \href{https://www.kaggle.com/datasets/waltervanhuissteden/the-nerthus-dataset}{Link} \\
    EndoSceneStill \cite{vazquez2017benchmark} & JHE'17 
    & 912 & 44 
    & & & \checkmark &
    & \#C1 $\rightarrow$ polyp 
    & \href{http://pages.cvc.uab.es/CVC-Colon/index.php/databases/cvc-endoscenestill/}{Link} \\ 
    \makecell[r]{KID1 \cite{koulaouzidis2017kid}\\ \\}
    &  \makecell[r]{EIO'17 \\ \\}
    & \makecell[r]{137\\ \\} 
    & \makecell[r]{-\\ \\}
    & \makecell[c]{\checkmark\\ \\} &  &  &
    & \makecell[l]{\#C10 $\rightarrow$ angiectasias, ulcers, stenoses, villous edema, nodular lymphangiectasias, chylous cysts, polyps, aphthae,\\normal/no pathology, intraluminal hemorrhage}
    & \makecell[l]{\href{https://mdss.uth.gr/datasets/endoscopy/kid/}{Link} \\ \\ } \\ 
    KID2 \cite{koulaouzidis2017kid} & EIO'17 
    & 2,371 & 47 
    & \checkmark & & \checkmark &
    & \#C4 $\rightarrow$ vascular anomalies, polypoid anomalies, inflammatory anomalies, normal images
    & \href{https://mdss.uth.gr/datasets/endoscopy/kid/}{Link} \\ 
    NBIPolyp-UCdb \cite{figueiredo2019unsupervised} & BSPC'19 
    & 86 & 11 
    &  &  & \checkmark  &
    & \#C2 $\rightarrow$ adenomas, hyperplastic
    & \href{http://www.mat.uc.pt/~isabelf/Polyp-UCdb/NBIPolyp-UCdb.html}{Link} \\ 
    WLPolyp-UCdb \cite{figueiredo2019polyp} & EIO'19 
    & 3,040 & 42 
    & \checkmark &  & \checkmark  &
    & \#C2 $\rightarrow$ polyp, normal mucosa 
    & \href{http://www.mat.uc.pt/~isabelf/Polyp-UCdb/WLPolyp-UCdb.html}{Link} \\ 
    ASEI \cite{hoang2019enhancing} & MM'19
    & 4,470 &-
    & \checkmark & \checkmark &  & 
    & \#C4 $\rightarrow$ dyed-lifted-polyps, dyed-resection-margins, instruments, polyp
    & \href{https://endoscopy.selab.hcmus.edu.vn/}{Link} \\
    Cho \etal \cite{cho2019identification} & PeerJ'19 
    & 328,927 & 112 
    & \checkmark &  &  &
    & \#C1 $\rightarrow$ cecum 
    & \href{https://figshare.com/articles/dataset/Colonoscopy_images/7937336/1}{Link} \\
    EAD2019 \cite{ali2019endoscopy} 
    & arXiv' 19
    & 2,342 & -
    &  & \checkmark & \checkmark &
    & \#C7 $\rightarrow$  imaging artefacts, contrast, specularity, instrument, bubbles, motion blur, saturation
    & \href{https://ead2019.grand-challenge.org/}{Link} \\
    Liu \etal \cite{liu2020photoshopping} & ISBI'20 
    & 14,317 & 18 
    & \checkmark &  &  &
    & \#C2 $\rightarrow$ polyp, non-polyp 
    & - \\
    Kvasir-SEG \cite{jha2020kvasir} & MMM'20 
    & 1,000 & - 
    & & \checkmark & \checkmark &
    & \#C1 $\rightarrow$ polyp 
    & \href{https://datasets.simula.no/kvasir-seg/}{Link} \\
    \makecell[r]{PICCOLO \cite{sanchez2020piccolo}\\ \\ \\} 
    & \makecell[r]{ApplSci'20\\ \\ \\}
    & \makecell[r]{3,433\\ \\ \\}
    & \makecell[r]{39 \\ \\ \\}
    & \makecell[c]{\checkmark\\ \\ \\}
    & 
    & \makecell[c]{\checkmark\\ \\ \\ } 
    &
    & \makecell[l]{\#C17 $\rightarrow$ Paris classification (protruded lesions: 0-Ip/0-Ips/0-Is, elevated lesions: 0-IIa/0-IIa+c, flat lesions: 0-IIb),\\NICE classification (type 1/2/3), Diagnosis (adenocarcinoma/adenoma/hyperplasia), Histological stratification\\(high grade dysplasia/hyperplasia/invasive adenocarcinoma/low grade dysplasia/no dysplasia)}
    & \makecell[c]{\href{https://www.biobancovasco.org/en/Sample-and-data-catalog/Databases/PD178-PICCOLO-EN.html}{Link}\\ \\ \\ } \\ 
    EDD2020 \cite{ali2020endoscopy}
    & arXiv'20
    & 386 
    & - 
    & \checkmark 
    & \checkmark 
    & \checkmark 
    &
    & \#C5 $\rightarrow$ suspicious area, high-grade dysplasia, adenocarcinoma, polyp, \tgray{normal dysplastic Barrett's oesophagus}
    & \href{https://edd2020.grand-challenge.org/}{Link} \\ 
    CAD-CAP \cite{leenhardt2020cad} & EIO'20 
    & 25,124 & 1,686 
    & \checkmark &  & \checkmark  &
    & \#C4 $\rightarrow$ vascular lesions, fresh blood, ulcero-inflammatory lesions, normal images
    & - \\
    \makecell[r]{ACP-ColonDB$_{530}$ \cite{poon2020ai}\\ \\ }
    & \makecell[r]{NPJDM'20 \\ \\ }
    & \makecell[r]{221,976\\ \\ }
    & \makecell[r]{-\\ \\ }
    & \makecell[c]{\checkmark\\ \\ } 
    & \makecell[c]{\checkmark\\ \\ }
    & 
    &
    & \makecell[l]{\#C13 $\rightarrow$ adenomatous polyp, hyperplastic polyp, other polyp, bleeding, IC valve, instrument, artefact, normal\\ colon structure, bubble, inside colon background, \tgray{stool, lumen, outside colon background}}
    & \makecell[c]{-\\ \\ } \\
    \makecell[r]{HyperKvasir~\cite{Borgli2020} \\ \\ \\}
    & \makecell[r]{SData'20 \\ \\ \\}
    & \makecell[r]{110,079 \\ \\ \\}
    & \makecell[r]{374 \\ \\ \\}
    & \makecell[c]{\checkmark \\ \\ \\}
    & \makecell[c]{\checkmark \\ \\ \\}
    & \makecell[c]{\checkmark \\ \\ \\}
    &
    & \makecell[l]{\#C23 $\rightarrow$ cecum, retroflex rectum, BBPS 0-1/2-3, ulcerative colitis grade 1/2/3/0-1/1-2/2-3, polyps, dyed lifted\\polyps, dyed resection margins, hemorrhoids, \tgray{Barrett's, terminal ileum, Z-line, esophagitis grade A, esophagitis}\\ \tgray{grade B-D, pylorus, retroflex stomach, Barrett's (short-segment), impacted stool}}
    & \makecell[c]{\href{https://datasets.simula.no/hyper-kvasir/}{Link} \\ \\ \\} \\
    WCE-Polyp \cite{guo2020learn} & TMI'20 
    & 541 & - 
    &  &  & \checkmark  &
    & \#C1 $\rightarrow$ polyp 
    & \href{https://github.com/Guo-Xiaoqing/ThresholdNet}{Link} \\
    EAD2020 \cite{ali2021deep} & MedIA'21
    & 2,531 & -
    &  & \checkmark & \checkmark &
    & \#C8 $\rightarrow$ specularity, bubbles, saturation, contrast, blood, instrument, blur, imaging artefacts
    & \href{https://ead2020.grand-challenge.org/}{Link} \\
    BKAI-Small \cite{ngoc2021neounet} & ISVC'21 
    & 1,200 & - 
    & & & \checkmark  &
    & \#C3 $\rightarrow$ non-neoplastic polyp, neoplastic polyp, background
    & \href{https://www.kaggle.com/c/bkai-igh-neopolyp/}{Link} \\
    BKAI-Large \cite{ngoc2021neounet} & ISVC'21 
    & 7,466 & - 
    & & &\checkmark  &
    & \#C4 $\rightarrow$ non-neoplastic polyp, neoplastic polyp, undefined polyp, background
    & \href{https://www.kaggle.com/c/bkai-igh-neopolyp/}{Link} \\ 
    CPC-Paired \cite{wang2021colorectal} & MICCAI'21 
    & 681 & - 
    & \checkmark & \checkmark &  &
    & \#C2 $\rightarrow$ hyperplastic polyp, adenoma
    & \href{https://github.com/qinwang-ai/PolypsAlign}{Link} \\
    LDPolyVideo \cite{ma2021ldpolypvideo} & MICCAI'21 
    & 901,666 & 263 
    & \checkmark & \checkmark & &
    & \#C2 $\rightarrow$ polyp, non-polyp 
    & \href{https://github.com/dashishi/LDPolypVideo-Benchmark}{Link}\\
    Celik \etal \cite{celik2021endouda} & MICCAI'21
    & 2,224 & -
    & \checkmark &  & \checkmark &
    & \#C2 $\rightarrow$ polyps, \tgray{Barrett’s esophagus} 
    & \href{https://github.com/sharib-vision/EndoUDA}{Link} \\
    Kvasir-Instrument \cite{jha2021kvasir} & MMM'21 
    & 590 & - 
    &  & \checkmark & \checkmark &
    & \#C1 $\rightarrow$ GI procedure tools (\eg, snares, balloons, and biopsy forceps)
    & \href{https://datasets.simula.no/kvasir-instrument/}{Link} \\
    CP-CHILD \cite{wang2020improved} & BMCMI'21 
    & 9,500 & - 
    & \checkmark & & &
    & \#C2 $\rightarrow$ colonic polyp, normal or other pathological images
    & \href{https://figshare.com/articles/dataset/CP-CHILD_zip/12554042}{Link} \\
    CROHN-IPI \cite{de2021multi} & EIO'21 
    & 3,498 & - 
    & \checkmark &  &  &
    & \#C7 $\rightarrow$ \makecell[l]{erythema, edema, aphthoid ulceration, ulceration (3–10mm, $>$10mm), stenosis, non-pathological} 
    & \href{https://crohnipi.ls2n.fr/en/crohn-ipi-project/}{Link} \\
    C-E Crohn’s Disease \cite{kong2021multi} & FMOLB'21 
    & 467 & 164 
    & \checkmark &  & \checkmark &
    & \#C1 $\rightarrow$ Crohn’s lesions
    & - \\
    \makecell[r]{SUN-database \cite{misawa2021development} \\ \\}
    & \makecell[r]{GIE'21  \\ \\} 
    & \makecell[r]{159,232  \\ \\} 
    & \makecell[r]{113  \\ \\} 
    & \makecell[c]{\checkmark  \\ \\} 
    & \makecell[c]{\checkmark  \\ \\} 
    &
    &
    & \makecell[l]{\#C7 $\rightarrow$ hyperplastic polyp, low grade adenoma, high-grade adenoma, traditional serrated adenoma, invasive\\carcinoma, sessile serrated lesion, negative}
    & \makecell[c]{\href{http://amed8k.sundatabase.org/}{Link} \\ \\}   \\ 
    Kvasir-Sessile \cite{jha2021comprehensive} & JBHI'21 
    & 196 & - 
    & & & \checkmark &
    & \#C1 $\rightarrow$ polyp ($<$10mm)
    & \href{https://datasets.simula.no/kvasir-seg/}{Link} \\
    \makecell[r]{Kvasir-Capsule \cite{smedsrud2021kvasir} \\ \\}
    & \makecell[r]{SData'21 \\ \\}
    & \makecell[r]{4,741,504 \\ \\}
    & \makecell[r]{117 \\ \\}
    & \makecell[c]{\checkmark \\ \\}
    & \makecell[c]{\checkmark \\ \\}
    &
    &
    & \makecell[l]{\#C14 $\rightarrow$ polyp, Ileocecal valve, lymphangiectasia, erythema, angiectasia, foreign body, erosion, ulcer, blood (fresh),\\blood (hematin), normal clean mucosa, reduced mucosal view, \tgray{pylorus, ampulla of Vater}}
    & \makecell[c]{\href{https://osf.io/dv2ag/}{Link} \\ \\} \\
    KUMC \cite{li2021colonoscopy} & PONE'21 
    & 37,899 & 155 
    & \checkmark & \checkmark &  &
    & \#C2 $\rightarrow$ hyperplastic polyps, adenomatous polyps
    & \href{https://dataverse.harvard.edu/dataset.xhtml?persistentId=doi:10.7910/DVN/FCBUOR}{Link}\\
    \makecell[r]{ERS* \cite{cychnerski2022ers}\\ \\ \\ \\}
    & \makecell[r]{arXiv'22 \\ \\ \\ \\ }
    & \makecell[r]{1,354,667 \\ \\ \\ \\ }
    & \makecell[r]{1,520 \\ \\ \\ \\ }
    & \makecell[c]{\checkmark \\ \\ \\ \\ }
    &  
    & \makecell[c]{\checkmark \\ \\ \\ \\ }
    &
    & \makecell[l]{\#C27 $\rightarrow$ ulcerative colitis (active/quiescent), stricture (postoperative/inflammatory/malignant), polyp, melanosis,\\diverticulosis, fistula, crohnsdisease (active/quiescent), lipoma, proctitis, hemorrhoids, submucosal tumor, solitary \\ulcer, bleeding of unknown origin, ileitis, diverticulitis, colitis: ischemic, colorectal cancer, angiodysplasia, rectal\\ulcer, foreign body, polyposis syndrome,  postoperative appearance, parasites}
    & \makecell[c]{\href{https://cvlab.eti.pg.gda.pl/en/publications/endoscopy-dataset}{Link} \\ \\ \\ \\ } \\
    Tian \etal \cite{tian2022contrastive} & MICCAI'22 
    & 807,069 & 253 
    & \checkmark & \checkmark &  &
    & \#C2 $\rightarrow$ polyp, non-polyp 
    & \href{https://github.com/tianyu0207/weakly-polyp}{Link}\\
    WCE-CCDD \cite{montalbo2022diagnosing} & BSPC'22 
    & 6,000 & - 
    & \checkmark &  &  &
    & \#C4 $\rightarrow$ ulcer, polyps, normal, \tgray{esophagitis}
    & \href{https://www.kaggle.com/datasets/francismon/curated-colon-dataset-for-deep-learning}{Link} \\
    PolypGen2.0 \cite{Ali2022EndoscopicCV} & ISBIw'22 
    & 3,446 & 46 
    & \checkmark & \checkmark & \checkmark &
    & \#C2 $\rightarrow$ serrated, adenomas
    & \href{https://endocv2022.grand-challenge.org/EndoCV-Sequence/}{Link} \\
    \makecell[r]{SUN-SEG \cite{ji2022video} \\ \\}
    & \makecell[r]{MIR'22 \\ \\}
    & \makecell[r]{159,232 \\ \\}
    & \makecell[r]{1,013 \\ \\}
    & \makecell[c]{\checkmark \\ \\}
    & \makecell[c]{\checkmark \\ \\}
    & \makecell[c]{\checkmark \\ \\}
    &
    & \makecell[l]{\#C7 $\rightarrow$ hyperplastic polyp, low grade adenoma, high-grade adenoma, traditional serrated adenoma, invasive\\carcinoma, sessile serrated lesion, negative}
    & \makecell[c]{\href{https://github.com/GewelsJI/VPS}{Link} \\ \\}\\
    SinGAN-Seg \cite{thambawita2022singan} & PONE'22 
    & 10,000 & - 
    & & & \checkmark &
    & \#C1 $\rightarrow$ polyp
    &\href{https://osf.io/xrgz8/}{Link}\\
    ENDOTEST \cite{fitting2022video} & SJG'22 
    & 253,754 & 58 
    & \checkmark & \checkmark &  &
    & \#C2 $\rightarrow$ polyp, non-polyp
    & \href{https://www.ukw.de/research/inexen/applied-ai/ai-for-polyp-detection/}{Link} \\
    MEDVQA-GI \cite{hicks2023overview} & CLEF'23 
    & 3,949 & - 
    & & & \checkmark & \checkmark
    & \#C2 $\rightarrow$ polyp, surgical equipment
    & \href{https://www.imageclef.org/2023/medical/vqa}{Link}\\
    \makecell[r]{GastroVision \cite{jha2023gastrovision}\\ \\ \\ \\ \\} 
    & \makecell[r]{ICMLw'23\\ \\ \\ \\ \\}
    & \makecell[r]{8,000 \\ \\ \\ \\ \\ }
    & \makecell[r]{-\\ \\ \\ \\ \\ }
    & \makecell[c]{\checkmark \\ \\ \\ \\ \\}
    & & &
    & \makecell[l]{\#C27 $\rightarrow$ accessory tools, angiectasia, blood in lumen, cecum, colon diverticula, resection margins, colorectal cancer,\\
    dyed-lifted-polyps, erythema, ulcer, dyed-resection-margins, retroflex rectum, mucosal inflammation large bowel,\\
    resected polyps, colon polyps, lleocecal valve, normal mucosa and vascular pattern in the large bowel, \tgray{esophagitis,}\\
    \tgray{Barrett's esophagus, duodenal bulb, esophageal varices, gastric polyps, gastroesophageal junction normal z-line,} \\
    \tgray{normal esophagus, normal stomach, pylorus, small bowel terminal ileum}}
    & \makecell[c]{\href{https://osf.io/84e7f/}{Link} \\ \\ \\ \\ \\} \\
    W-Polyp \cite{ren2023towards} & CVPR'23 
    & 1,450 & - 
    & & & \checkmark &
    & \#C1 $\rightarrow$ polyp 
    & \href{https://github.com/ic-qialanqian/WS-DefSegNet}{Link}\\
    LIMUC \cite{polat2023improving} & IBD'23
    & 11,276 & - 
    & \checkmark &  &  &
    & \#C4 $\rightarrow$ Mayo endoscopic score (MES) 0/1/2/3
    & \href{https://zenodo.org/records/5827695#.ZF-92OzMJqs}{Link} \\
    PS-NBI2K \cite{psnbi2k} & JBHI'23 
    & 2,000 & - 
    & & & \checkmark &
    & \#C1 $\rightarrow$ polyp 
    & \href{https://github.com/JaeZ1205/PS_NBI2k}{Link} \\ 
    PolypGen \cite{ali2021polypgen} & SData'23 
    & 8,037 & 23 
    & \checkmark & \checkmark & \checkmark &
    & \#C2 $\rightarrow$ polyp, negative
    & \href{https://www.synapse.org/Synapse:syn26376615/wiki/613312}{Link}\\
    MedFMC** \cite{wang2023real} & SData'23     
    & 22,349 & -
    & \checkmark &  &  &
    & \#C5 $\rightarrow$ ulcer, erosion, polyp, tumor, and non-instance
    & \href{https://doi.org/10.6084/m9.figshare.c.6476047.v1}{Link}\\
    GB-WCE Dataset \cite{GBWCEDataset} &  MD'23
    & 226 & -
    & \checkmark &  &  &
    & \#C2 $\rightarrow$ bleeding or lesions, normal
    & \href{https://data.mendeley.com/datasets/8pbbjf274w/1#:~:text=The%20dataset%20contains%20the%20wireless}{Link} \\
    REAL-Colon \cite{biffi2024real} & SData'24 
    & 2,757,723 & 60
    & & \checkmark & &
    & \#C2 $\rightarrow$ polyp, negative 
    & \href{https://plus.figshare.com/articles/media/REAL-colon_dataset/22202866}{Link} \\
    Xu \etal \cite{xu2024ssl} & TMI'24 
    & 251 & -
    & \checkmark &  &  &
    & \#C4 $\rightarrow$ Mayo endoscopic score (MES) 0/1/2/3
    & \href{https://www.synapse.org/Synapse:syn52674005/files/}{Link} \\
    Kvasir-VQA \cite{gautam2024kvasirvqa} & MMw'24
    & 6,500 & -
    & \checkmark &  &  & \checkmark
    & \#C5 $\rightarrow$ polyps, ulcerative colitis, instrument, normal, \tgray{esophagitis}
    & \href{https://datasets.simula.no/kvasir-vqa/}{Link} \\
%
    CapsuleVision2024 \cite{handa2024capsule}
    & CVIP'24
    & 58,124
    & -
    & \checkmark &  &  &
    & \#C10 $\rightarrow$ angioectasia, bleeding, erosion, erythema, foreign body, lymphangiectasia, polyp, ulcer, worms, normal
    & \href{https://github.com/UTSAVS26/Capsule-Vision-2024-Challenge}{Link} \\
    COLON \cite{ruiz2024colon} & arXiv'24 
    & $\sim$430,000 & 30
    & \checkmark & \checkmark & \checkmark &
    & \#C3 $\rightarrow$ adenoma, hyperplastic, non-pathological case
    & - \\ 
    WCEBleedGen \cite{handa2024wcebleedgen} & arXiv'24
    & 2,618 & -
    & \checkmark & \checkmark & \checkmark &
    & \#C2 $\rightarrow$  bleeding, non-bleeding
    & \href{https://zenodo.org/records/10156571}{Link} \\
    PolypDB \cite{jha2024polypdb} & arXiv'24 
    & 3,934 & -
    &  & \checkmark & \checkmark &
    & \#C1 $\rightarrow$ polyp (multiple imaging modalities and multiple medical centers)
    & \href{https://github.com/DebeshJha/PolypDB}{Link} \\
%
\end{tabular}
\begin{tablenotes}
\item \textit{*NOTE} -- The ERS dataset \cite{cychnerski2022ers} includes 99 annotated categories in total. For the sake of brevity, we list only 27 colon-related categories within ERS.
\item \textit{**NOTE} -- The MedFMC dataset \cite{wang2023real} comprises 23,349 medical images across five modalities. This table only enumerates the categories specific to the endoscopic modality.
\end{tablenotes}
\end{table*}

\section{Background}\label{sec:background}

\subsection{Origin and evolution}\label{sec:colonoscopy_origins_and_evolution}
The history of colonoscopy has two key milestones. In 1968, gastrointestinal surgeons Hiromi Shinya and William Wolff found a link between colonic polyps and intestinal tumours, but they lacked equipment to examine them. In 1969, they discovered Corning Incorporated's optical fibres and collaborated with Olympus to create the fiberoptic colonoscope, a groundbreaking device to examine the colon and remove polyps using wire loops. The second milestone came in 1983 with the introduction of the electronic colonoscope \cite{sivak2006gastrointestinal}, which allows visualisation of the colon on a screen and polyp removal using a polypectomy snare, enhancing detection rates and reducing bleeding. The 21st century brings the AI era, where computer-aided diagnosis systems provide greater precision and reliability in procedures~\cite{berzin2020adding}.
This study explores the transformative impact of intelligent techniques for colonoscopy, a type of endoscopy \cite{iddan2000wireless}, while other related areas such as laparoscopy \cite{twinanda2016endonet} are discussed in \pink{appendix}.

\subsection{Intrinsic traits and domain-unique challenges}\label{sec:domain_unique_challenges}

We summarise five unique challenges associated with colonoscopic vision tasks, primarily caused by procedural aspects and imaging conditions during a colonoscopy.
\textit{(a) Non-linear camera ego-motion.} Procedural constraints force the camera (\ie, colonoscope) to actively move in a non-linear and unpredictable manner, challenging ego-motion compensation \cite{shao2022self} and causing motion blur \cite{ji2022video}.
\textit{(b) Presence of medical instruments.} The colonoscopy procedure often includes instruments such as scopes, guidewires, and snares, which should be distinguished properly from anatomical structures \cite{ceron2022real} for efficient analysis.
\textit{(c) Limited observable field.} The intricate folds and blind spots within the colon restrict the visible area in colonoscopy data. This requires algorithms capable of extracting relevant information from limited visual landscapes \cite{blau2021unsupervised}.
\textit{(d) Non-uniform illumination.} The mucosal surface of the colon, prone to wetness and sheen, results in highly variable and diffuse illumination with complex reflections such as non-Lambertian reflections and interreflections. Traditional lighting-based algorithms struggle under these conditions \cite{zhang2021lighting}.
\textit{(e) Variability in tissue appearance.} Mucosal textures and colours vary considerably due to constant movement, disease states, anatomical differences, and instrument effects. Furthermore, benign polyps or lesions usually have weak or homogeneous boundaries \cite{fan2023advances}, making them blend into surrounding tissues and difficult to detect. These issues require a robust response from AI models to inherent morphological and colour fluctuations.

\section{Revisiting Colonoscopy Data}\label{sec:colonoscopy_related_datasets}

\subsection{Medical data for colonoscopy}\label{sec:colonoscopy_dataset_revisit}


\tabref{tab:colonoscopy_datasets} presents our investigation of {\RevDataNum} datasets with their essential statistics for four tasks related to colonoscopic scene perception.
We search for them using queries such as ``colonoscopy dataset/benchmark'' and ``gastrointestinal disease dataset''. They consist of images or videos related to the human colon. 
In particular, some datasets also include images of other organs, such as the pylorus in \cite{pogorelov2017kvasir} or the stomach in \cite{jha2023gastrovision}. 
Next, we review these selected datasets according to their different task objectives.


\noindent$\bullet$~\textbf{Classification datasets}
have been widely used for varied purposes, such as colon disease classification in images \cite{de2021multi,montalbo2022diagnosing,polat2023improving,xu2024ssl,pogorelov2017kvasir,ali2020endoscopy,jha2023gastrovision,Borgli2020,koulaouzidis2017kid,cychnerski2022ers,handa2024capsule}/videos \cite{misawa2021development,ji2022video,smedsrud2021kvasir}, 
polyp identification \cite{angermann2017towards,wang2020improved,figueiredo2019polyp,ma2021ldpolypvideo,liu2020photoshopping,leenhardt2020cad,sanchez2020piccolo}, fine-grained polyp classification \cite{zhang2016automatic,wang2021colorectal,mesejo2016computer}, bleeding condition \cite{deeba2016automated,handa2024wcebleedgen,GBWCEDataset}, anomaly recognition \cite{tian2022contrastive}, cecum recognition \cite{cho2019identification}, and pre-operative assessment \cite{pogorelov2017nerthus}.

\noindent$\bullet$~\textbf{Detection datasets} provide both categorical and localisation labels for targets of interest,
such as colonic diseases \cite{misawa2021development,li2021colonoscopy,ji2022video}, accessory instruments \cite{poon2020ai,hoang2019enhancing,poon2020ai}, polyps \cite{ma2021ldpolypvideo,tian2022contrastive,ye2016online,fitting2022video,biffi2024real}, endoscopic artefacts \cite{ali2021deep,ali2019endoscopy}, and gastrointestinal diseases \cite{smedsrud2021kvasir,Borgli2020}. 
In addition, the organisation of competitions has accelerated growth within the colonoscopy community by establishing shared platforms for data collection and model evaluation, greatly advancing research areas like the detection of intestinal disease \cite{ali2020endoscopy} and polyp \cite{Ali2022EndoscopicCV,ali2021polypgen,jha2020kvasir}.

\noindent$\bullet$~\textbf{Segmentation datasets} for colonoscopy research originate from two sources. The first source comprises real data, mainly utilised for single-target segmentation of entities such as polyps \cite{psnbi2k,vazquez2017benchmark,figueiredo2019unsupervised,figueiredo2019polyp,guo2020learn,celik2021endouda}, Crohn's disease \cite{kong2021multi}, and accessory tools \cite{jha2021kvasir}. Some datasets, like BKAI-Small/Large \cite{ngoc2021neounet}, provide instance-level masks for neoplastic and non-neoplastic polyps. Other datasets come from organised competitions, such as polyp segmentation datasets \cite{bernal2012towards,silva2014toward,bernal2015wm,tajbakhsh2015automated,Ali2022EndoscopicCV} and a gastrointestinal diseases segmentation dataset \cite{ali2020endoscopy}, or extensions of existing databases,  offering pixel-wise masks (\eg, for polyps \cite{jha2020kvasir,jha2021comprehensive}, colorectal disease \cite{ji2022video}) or scribble labels (\eg, for polyps \cite{tian2022contrastive,ren2023towards}). The other source, such as SinGAN-Seg \cite{thambawita2022singan}, generates synthetic images for polyp segmentation.

\noindent$\bullet$~\textbf{VL datasets} remain relatively scarce so far, with two known datasets for this specific purpose. MEDVQA-GI \cite{hicks2023overview} contributes the first dataset with three VL tasks, including visual question answering, visual question generation, and visual location question answering. Kvasir-VQA \cite{gautam2024kvasirvqa} collects 6,500 question-answer pairs from existing datasets \cite{Borgli2020,jha2021kvasir}, for gastrointestinal diagnostics, such as image captioning, visual question answering.



%
\subsection{Discussion}\label{sec:colonoscopy_dataset_discussion}

Based on the {\RevDataNum} revisited datasets, we have several data-centric observations that could inspire more future ideas.


\noindent$\bullet$~\textbf{Data granularity} requires improvement to better understand patient conditions and treatment efficacy. 
\textit{(a)} More than a quarter of polyp-containing datasets provide fine-grained categorisation, often with inadequate detail. For example, BKAI-Small/-Large \cite{ngoc2021neounet} provide two instance annotations of neoplastic and non-neoplastic polyps. ColonoscopicDS \cite{mesejo2016computer} categorises at the video level into hyperplasic, serrated, and adenoma lesions. SUN-database \cite{misawa2021development} provides fine-grained labels, documenting measurements of polyp size (height, width) and morphology (pedunculated, sessile, flat), along with their anatomical locations (\eg, rectum, sigmoid colon). Several data-centric areas remain under-explored, such as temporal lesion evolution recording, granularity improvement, graded severity tagging, and instance-level target annotation. 
\textit{(b)} Label orthogonality, an often overlooked issue, treats classes as isolated entities. Current works seldom discuss potential inter-class correlations, such as the co-occurrence of inflammatory bowel disease with erosion symptoms, Crohn's disease with fistula complications, or colorectal cancer accompanied by bleeding. Future studies should consider causality \cite{de2012role} and comorbidity \cite{li2009pathway} to address these correlations effectively.

\noindent$\bullet$~\textbf{Data diversity} is crucial to developing fair and reliable models. Three aspects deserve consideration.
\textit{(a)} Datasets for rare colorectal diseases appear to be limited, due to case scarcity and expertise requirements. For example, Crohn's disease, which affects an estimated 58 to 241 per 100,000 adults in the United States \cite{veauthier2018crohn}, has so far been discussed in three datasets \cite{kong2021multi,cychnerski2022ers,de2021multi}.
Such an unbalanced situation leads to data-hungry models performing better in common cases than in rarer or novel ones. Thus, increasing attention to rare gastrointestinal diseases could potentially improve the ability to treat long-tailed \cite{yang2022survey} or open-vocabulary \cite{wu2024towards} problems.
\textit{(b)} Multimodal research in colonoscopy appears to be in its early stages, with limited data \cite{hicks2023overview,gautam2024kvasirvqa} for analysis. Therefore, collecting varied patient information (\eg, age, gender, eating habits) combined with expert interpretations (\eg, clinical report, medication advice) could be advantageous and ultimately facilitate personalised and side-effect-minimised colonoscopy practices \cite{singhal2023large}. 

\noindent$\bullet$~\textbf{Data inconsistency} in colonoscopy research is due to two main factors. 
\textit{(a)} Expert interpretations vary due to varying experience, expertise, and observed nuances, leading to subjective judgements and labelling uncertainties. For example, ColonoscopicDS \cite{mesejo2016computer} provides raw diagnostic labels for each sample from multiple experts and beginners, reflecting their underlying (dis)agreement. In addition, SUN-SEG \cite{ji2022video} releases the rejected segmentation masks from their annotation workflow, highlighting the challenges in reaching consistent polyp boundaries.
\textit{(b)} Existing datasets often have study-specific targets, leaving others unlabelled or classified as background. Nerthus \cite{pogorelov2017kvasir}, for example, focuses on assessing the quality of bowel preparation, but ignores diagnostic findings such as polyps. Furthermore, multiple categories in GastroVision \cite{jha2023gastrovision} are not mutually exclusive; for example, a case labelled as ``accessory tool'' could also fall into the category ``blood in lumen''.
Segmentation data like Kvasir-Instrument \cite{jha2021kvasir} annotates only medical instruments, ignoring other targets like polyps, whereas Kvasir-SEG \cite{jha2020kvasir} provides polyp label, leaving out others like instruments. These observations prompt future research into learning from partial \cite{zhang2021dodnet}, noisy \cite{karimi2020deep}, or missing \cite{yu2014large} labels.

\begin{table*}[t!]
\centering
\fontsize{6.2pt}{7pt}\selectfont
\renewcommand{\arraystretch}{0.8}
\renewcommand{\tabcolsep}{1.55mm}
\caption{\textbf{Summary of classification models in colonoscopy.}
\underline{Dataset:} 
CU=CU-ColonDB \cite{zhang2016automatic}, 
CDS=ColonoscopicDS \cite{mesejo2016computer},
Private= private data,
HK=HyperKvasir \cite{Borgli2020}, 
KC=Kvasir-Capsule \cite{smedsrud2021kvasir}.
%
\underline{Backbone:} 
CaffeNet \cite{jia2014caffe}, 
D-121=DenseNet121 \cite{huang2017densely}, 
R-12/-18/-50/ -101=ResNet12/18/50/101 \cite{he2016deep},
ViT-S16/ViT-B16 \cite{dosovitskiy2021vit}, 
MobV2=MobileNetV2 \cite{sandler2018mobilenetv2}, 
R50-Att=ResNet50 with attention module \cite{woo2018cbam}, 
C3D \cite{tran2015learning}, 
Inc-v3=Inceptionv3 \cite{szegedy2016rethinking}, 
I3D \cite{carreira2017quo}. 
``Customised'' means a base network modified for the current task or a model independent of the base network choice.
\underline{Head:} classifier implemented by the fully connected (FC) and support vector machine (SVM) layers, or using the $\ell^2$ norm to measure the disparity between the input and output.
\underline{Arch:} the architectures shown in \figref{fig:architecture_taxonomy}. 
\underline{Sup:} learning strategies such as fully supervised (FS), semi-supervised (SS), unsupervised (US), and weakly supervised (WS). For simplicity, the following tables use consistent abbreviations unless specified otherwise.
}
\label{tab:classification_models}
\begin{tabular}{c|r|r|c|c c|c c|c c|c}
    & Model & Publication & Core design &Training dataset &Testing dataset &Backbone &Arch &Head  &Sup  &URL \\
\hline
    \multirow{20}{*}{\begin{sideways}Image-based models\end{sideways}}
    & Zhang~\etal~\cite{zhang2016automatic} & JBHI'16 & domain transfer learning & CU, CDS & CU, CDS & CaffeNet & BF\#1 & SVM & FS & - \\
    & RIIS-DenseNet~\cite{yuan2018riis} & MICCAI'18 & rotation-invariant, similarity constrained & Private & Private & D-121 & SF & FC & FS & - \\
    & FSAD-Net~\cite{tian2020few} & MICCAI'20 & mutual information maximisation & Private & Private & D-121 & BF\#2 & FC & US & \href{https://github.com/tianyu0207/FSAD-Net}{Link} \\
    & Gammulle~\etal~\cite{gammulle2020two} & MICCAI'20 & relational mapping & Kvasir \cite{pogorelov2017kvasir}, Nerthus \cite{pogorelov2017nerthus} & Kvasir \cite{pogorelov2017kvasir}, Nerthus \cite{pogorelov2017nerthus} & R-50 & MF\#1 & FC & FS & - \\
    & ADGAN~\cite{liu2020photoshopping} & ISBI'20 & dual adversarial learning & Liu~\etal \cite{liu2020photoshopping} & Liu~\etal \cite{liu2020photoshopping} & Customised & BF\#2 & $\ell^2$ & US & - \\
    & Carneiro~\etal~\cite{carneiro2020deep} & MedIA'20 & model uncertainty \& calibration & Private & Private &D-121 & SF & FC & FS & - \\ 
    & SSL-WCE~\cite{guo2020semi} & MedIA'20 & adaptive aggregated attention & CAD-CAP \cite{leenhardt2020cad} &  CAD-CAP \cite{leenhardt2020cad} & D-121 & BF\#2 & FC & SS & \href{https://github.com/Guo-Xiaoqing/SSL_WCE}{Link} \\
    & PolypsAlign~\cite{wang2021colorectal} & MICCAI'21 & teacher-student alignment & CPC-Paired \cite{wang2021colorectal} & CPC-Paired \cite{wang2021colorectal} & R-50 & BF\#2 & FC & FS & \href{https://github.com/qinwang-ai/PolypsAlign}{Link} \\
    & CPC-Trans~\cite{ma2022toward} & MICCAI'22 & cross-modal representation consistency & CPC-Paired \cite{wang2021colorectal} & CPC-Paired \cite{wang2021colorectal} & ViT-S16 & BF\#2 & FC & FS & \href{https://github.com/WeijieMax/CPC-Trans}{Link} \\
    & FFCNet~\cite{wang2022ffcnet} & MICCAI'22 & frequency domain learning & Private & Private & R-18 & SF & FC & FS & \href{https://github.com/soleilssss/FFCNet}{Link} \\
    & DLGNet~\cite{wang2023dlgnet} & MedIA'23 & Gaussian mixture model & Private & Private & R-18 & BF\#2 & FC & FS & \href{https://github.com/soleilssss/DLGNet}{Link} \\
    & Yue~\etal~\cite{yue2023automated} & TIM'23 & class imbalance loss & Private, HK & Private, HK & MobV2 & SF & FC & FS & \href{https://github.com/Weipeishan2021/Class_Imbalance_loss}{Link} \\
    & DAFON~\cite{luo2024dynamic} & ESWA'24 & few-shot open-set learning & Kvasir-Capsule \cite{smedsrud2021kvasir} & Kvasir-Capsule \cite{smedsrud2021kvasir} & R-12 & BF\#2 & FC & FS & - \\
    & SSL-CPCD~\cite{xu2024ssl} & TMI'24 & composite pretext-class discrimination & LIMUC \cite{polat2023improving} & Private, LIMUC \cite{polat2023improving} & R50-Att & BF\#2 & FC & FS & \href{https://github.com/EricXuziang/SSL-CPCD}{Link} \\
%
%
\hline
    \multirow{6}{*}{\begin{sideways}Video\end{sideways}}
    & BseNet~\cite{itoh2018towards} & MICCAI'18 & unsupervised depth estimation, LSTM~\cite{6795963} & Private & Private & C3D & SF & FC & FS & - \\
    & Byrne~\etal~\cite{byrne2019real} & Gut'19 & real-time assessment system & Private & Private & Inc-v3 & SF & FC & FS & - \\
    & Tamhane~\etal~\cite{tamhane2022colonoscopy} & MICCAIw'22 & vision transformer based & Private & Private & ViT-B16 & SF & FC & FS & - \\
    & Tian~\etal~\cite{tian2022contrastive} & MICCAI'22 & multiple instance learning & WVAD \cite{tian2022contrastive} & WVAD \cite{tian2022contrastive}  & I3D & SF & FC & WS & \href{https://github.com/tianyu0207/weakly-polyp}{Link} \\
%
\end{tabular}
\end{table*}

\begin{figure*}[t!]
    \centering
    \includegraphics[width=\textwidth]{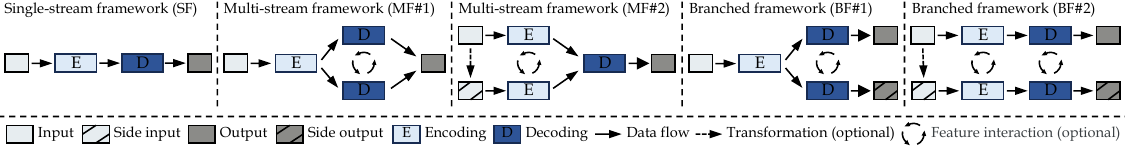}
    \vspace{-15pt}
    \caption{\textbf{Gallery of deep-based architectures.} The single-stream framework (SF) features a single input and output with sequential data flow. Multi-stream frameworks predict a single output but involve parallel processing streams, either at the decoding stage (MF\#1) or the encoding stage (MF\#2). Branched frameworks extend multi-stream framework to produce multiple outputs from either a single input (BF\#1) or multiple inputs (BF\#2). These side outputs typically receive supervision from additional supervisory signals, such as boundary cues.}
    \label{fig:architecture_taxonomy}
\end{figure*}

\section{Revisiting Colonoscopy Models}
\label{sec:colonoscopy_models}

In this section, we investigate {\RevPaperNum} deep learning models for colonoscopic scene perception, sourced from leading conferences or journals published since 2015. First, three widely recognised topics are described, including 18 classification models (\secref{sec:classification_models}), 24 detection models (\secref{sec:detection_models}), and 86 segmentation models (\secref{sec:segmentation_models}). Their architectural designs are classified into five subtypes as presented in \figref{fig:architecture_taxonomy}. Lastly, we discuss nine VL-related models in \secref{sec:vision_language_models}.

\subsection{Classification models}
\label{sec:classification_models}

\noindent$\bullet$~\textbf{Input phase.} \tabref{tab:classification_models} lists the training and testing data used for each deep model.
We note that many classification models \cite{yuan2018riis,carneiro2020deep,wang2022ffcnet,yue2023automated,wang2023dlgnet,itoh2018towards,byrne2019real} in colonoscopy use in-house data for model development, partly resulting in the absence of domain-recognised benchmarks. This issue stems from the different categorical goals pursued by individual studies, such as identifying polyps from white-light and narrow-band imaging pairs \cite{wang2021colorectal,ma2022toward}, images \cite{yuan2018riis}, or videos \cite{tian2022contrastive}, evaluating polyp size \cite{itoh2018towards}, and recognising colonic diseases \cite{wang2022ffcnet} or landmarks \cite{tamhane2022colonoscopy}.

\noindent$\bullet$ \textbf{Processing phase.} We discuss data flow management strategies based on two attributes.
%
\textit{(a) Backbone:} Early models \cite{gammulle2020two,yuan2018riis} typically employ well-trained convolutional backbones (\eg, ResNet \cite{he2016deep}, DenseNet \cite{huang2017densely}) from ImageNet \cite{deng2009imagenet}, while recent studies explore alternatives such as using a vision transformer in \cite{ma2022toward} and a lightweight network in \cite{yue2023automated}. Another strategy is SSL-CPCD \cite{xu2024ssl}, which involves pre-training a model to generate domain-specific representations, followed by its generalisation on various downstream perception tasks.
\textit{(b) Architecture:} Classification models involve various designs as in \figref{fig:architecture_taxonomy}. At first, a basic idea is of using a single-stream framework (SF) that sequentially processes visual features based on confidence calibrated \cite{carneiro2020deep} or 3D convolutional \cite{itoh2018towards,tian2022contrastive} networks. Second, Gammulle \etal \cite{gammulle2020two} proposed a dual decoding flow approach for hierarchical feature encoding, typified as MF\#1. 
Third, to ensure reliable predictions, branched frameworks are adopted for multi-objective learning, such as the integration of {parallel feature flows \cite{zhang2016automatic},} interclass Gaussian loss \cite{wang2023dlgnet}, and global-to-local consistency \cite{ma2022toward}.

\begin{table*}[t!]
\centering
\fontsize{6.2pt}{7pt}\selectfont
\renewcommand{\arraystretch}{0.8}
\renewcommand{\tabcolsep}{0.7mm}
\caption{\textbf{Summary of detection models in colonoscopy.} 
\underline{Dataset:}
C6=CVC-ClinicDB \cite{bernal2015wm},
ETIS=ETIS-Larib \cite{silva2014toward}, 
ASEI \cite{hoang2019enhancing}, 
C3=CVC -ColonDB \cite{bernal2012towards},
KUMC \cite{li2021colonoscopy},
LDPV=LDPolyVideo \cite{ma2021ldpolypvideo},
SUN=SUN-database \cite{misawa2021development},
PL=PICCOLO \cite{sanchez2020piccolo}, 
KID=KID1\&2 \cite{koulaouzidis2017kid},
CDS \cite{mesejo2016computer}, 
KSe=Kvasir-Sessile \cite{jha2021comprehensive},
ASU=ASU-Mayo \cite{tajbakhsh2015automated}, 
CDB=CVC-ClinicVideoDB \cite{angermann2017towards},
ES=EndoSceneStill \cite{vazquez2017benchmark}, 
CU \cite{zhang2016automatic}, 
ACP= ACP-ColonDB530 \cite{poon2020ai}.
\underline{Backbone:} 
R-34/-50/-101 \cite{he2016deep}, 
CDN-53=CSPDarkNet53 \cite{bochkovskiy2020yolov4},
DN-53=DarkNet-53 \cite{redmon2018yolov3}, 
EffDet-D0= EfficientDet-D0 \cite{tan2020efficientdet}, 
AlexNet \cite{li2019siamrpn++}, 
V-16=VGG16 \cite{simonyan2014very}, 
R-50v2=ResNet50V2 \cite{he2016identity}.
\underline{WF:} one-stage (OS) or two-stage (TS) workflows. 
%
%
\underline{NMS:} non-maximum suppression.
\underline{EC:} edge-sensitive cues. 
%
}
\label{tab:detection_models}
\begin{tabular}{c|r|r|c|c c|c c c|c c c|c}
    & Model & Publication &Core design &Training dataset &Testing dataset &Backbone &WF &Arch &NMS &EC &Sup  & URL \\
\hline
\multirow{18}{*}{\begin{sideways}Image-based models\end{sideways}}
%
    & Yang \etal~\cite{yang2020colon} & TIM'20 & parallel detection \& segmentation & Private, C6, ETIS & Private, C6, ETIS & R-50 & TS & BF\#1 &  & & FS & - \\
    & ConsolidatedPolypDA~\cite{liu2021consolidated} & MedIA'21 & Gaussian Fourier domain adaptation & C6 & ETIS, ASEI & R-101 & TS & BF\#2 &  &  & US & \href{https://github.com/CityU-AIM-Group/ConsolidatedPolypDA}{Link} \\
    & MDeNetplus~\cite{qadir2021toward} & MedIA'21 & 2D Gaussian shapes prediction & C6 & C3, ETIS & R-34 & OS & MF\#1 &  & \checkmark & FS & - \\
    & FedInI \cite{liu2022intervention} & MICCAI'22 & federated learning, structural causal model  & KUMC & KUMC & R-101 & TS & BF\#2 & \checkmark &  & FS & \href{https://github.com/CityU-AIM-Group/FedInI}{Link}  \\
    & Pacal~\etal~\cite{pacal2022efficient} & CIBM'22 & improved YOLOv3~\cite{redmon2018yolov3}/v4~\cite{bochkovskiy2020yolov4} & SUN, PL & SUN, PL, ETIS & CDN-53/DN-53 & OS & BF\#1 & \checkmark &  & FS & \href{https://github.com/ishakpacal/An-Efficient-Real-Time-Colonic-Polyp-Detection-with-YOLO-Algorithms-Trained-by-Using-Negative-Sample}{Link} \\
    & SMPT++~\cite{liu2022source} & TMI'22 & source-free domain adaptation & \makecell{Private, C6,\\ETIS, ASEI, KID} & \makecell{Private, C6,\\ ETIS, ASEI, KID} & R-101 & OS & BF\#1 & \checkmark &  & US & \href{https://github.com/CityU-AIM-Group/SFPolypDA}{Link} \\
%
%
    & FRCNN-AA-CIF~\cite{gong2023frcnn} & CIBM'23 & attention module \& context information fusion & Private & Private & R-101 & TS & BF\#1 & \checkmark &  & FS & - \\
%
%
    & Haugland~\etal~\cite{haugland2023deep} & MI'23 & modality translation  & Private, PL, CDS & PL, KUMC & EffDet-D0 & OS & BF\#2 & \checkmark &  & FS & - \\
    & SCAN++~\cite{li2023scan++} & TMM'23 & enhanced semantic conditioned adaptation & C6, ASEI & C6, ASEI & R-101 & OS & BF\#2 &  &  & FS, US & \href{https://github.com/CityU-AIM-Group/SCAN/tree/SCAN++}{Link} \\
    & TFCNet~\cite{pan2024tfcnet} & CIBM'24 & fine-grained feature compensation &  C6, KUMC, LDPV &  C6, KUMC, LDPV, KSe & CDN-53 & OS & BF\#1 &  &  & FS & - \\
    & DUT~\cite{liu2024decoupled} & TNNLS'24 & decoupled unbiased teacher & C6, ASEI, Private & ASEI, Private & R-101 & OS & BF\#2 &  &  & US & \href{https://github.com/CityU-AIM-Group/Decoupled-Unbiased-Teacher}{Link} \\
\hline
\multirow{20}{*}{\begin{sideways}Video-based models\end{sideways}}
    & Tajbakhsh~\etal~\cite{tajbakhsh2015comprehensive} & IPMI'15 & patch descriptor \& edge classification & Private & Private & AlexNet & TS & BF\#1 &  & \checkmark & FS & - \\
    & Tajbakhsh~\etal~\cite{tajbakhsh2015automated} & TMI'15 & extension on \cite{tajbakhsh2015comprehensive} & C3 & C3, ASU & AlexNet & TS & BF\#1 &  & \checkmark & FS & - \\
    & Yu~\etal~\cite{yu2016integrating} & JBHI'16 & online and offline integration & ASU & ASU & Customised & OS & MF\#2 &  &  & FS & - \\
%
%
    & Mo~\etal~\cite{mo2018efficient} & ICPR'18 & building upon Faster R-CNN~\cite{ren2015faster} & CDB & C6, C3, CDB, ES & V-16 & TS & BF\#1 & \checkmark &  & FS & - \\
    & Qadir~\etal~\cite{qadir2019improving} & JBHI'19 & temporal dependency & ASU, C6 & ASU, CDB & V-16 & TS & BF\#1 & \checkmark &  & FS & - \\
%
    & AIPDT~\cite{zhang2020asynchronous} & MICCAI'20 & parallel detection and tracking & Private, CDB & CDB & DN-53, AlexNet & OS & BF\#2 &  &  & FS & - \\
    & AI-doscopist~\cite{poon2020ai} & NPJDM'20 & spatial-temporal fusion & \makecell{C6, ETIS, C3,\\ASU, CU, ACP} & \makecell{C6, ETIS, C3,\\ASU, CU, ACP} & R-50 & OS & BF\#2 & \checkmark &  & FS & - \\
    & STFT~\cite{wu2021multi} & MICCAI'21 & spatial-temporal feature transformation & ASU, CDB & ASU, CDB & R-50 & OS & BF\#2 &  &  & FS & \href{https://github.com/lingyunwu14/STFT}{Link} \\
    & Yu~\etal~\cite{yu2022end} & AIM'22 & instance tracking head (plug-and-play) & Private, C6, CDB & Private, CDB, ETIS & V-16 & OS & BF\#2 & \checkmark &  & FS & - \\
    & EMSEN~\cite{wang2022explainable} & TII'22 & explainable multitask Shapley explanation & CDS & CDS & Customised & OS & BF\#2 &  &  & FS & - \\
    & YONA~\cite{jiang2023yona} & MICCAI'23 & feature alignment \& contrastive learning & SUN, CDB, LDPV & SUN, CDB, LDPV & R-50 & TS & BF\#2 & \checkmark &  & FS & \href{https://github.com/yuncheng97/YONA}{Link} \\
    & Intrator~\etal~\cite{intrator2023self} & MICCAI'23 & self-supervised polyp re-identification & Private & Private & R-50v2 & OS & MF\#2 &  &  & US & - \\
    & V2I-DETR~\cite{jiang2024let} & BIBM'24 & video-to-image knowledge distillation & SUN & SUN & R-50 & OS & BF\#2 &  &  & FS & - \\
%
%
\end{tabular}
\end{table*}

\noindent$\bullet$~\textbf{Output phase.} 
%
\textit{(a) Prediction head:} 
An early model~\cite{zhang2016automatic} applies two SVM layers to classify polyps into three categories. Modern methods usually adopt a fully connected layer as the final classifier due to its simplicity and flexibility. A special case is ADGAN \cite{liu2020photoshopping}, a generative adversarial network that computes the $\ell^2$ norm differential between input and output to identify anomalous images.
%
\textit{(b) Supervision strategy:} 
Most methods use fully-supervised learning with pre-annotated categories, but some explore data-efficient strategies, including semi-supervised~\cite{guo2020semi}, weakly-supervised~\cite{tian2022contrastive}, and unsupervised~\cite{liu2020photoshopping,tian2020few} learning.

%
\noindent$\bullet$~\textbf{Remarks.} We observe three aspects of the above classification models. \textit{(a)} Novel visual backbones, like the state space model \cite{zhu2024vision}, remain underexplored. Furthermore, reformulating the classification paradigm within VL models, \eg, CLIP \cite{radford2021clip}, can yield unexpected results.
\textit{(b)} Benchmarks for multicategory classification remain underexplored. The Kvasir series \cite{pogorelov2017kvasir,Borgli2020,smedsrud2021kvasir} offers valuable sources for further study.
We will explore these public data on their potential synergy in \secref{sec:colon300k}.
\textit{(c)} Several new task settings have emerged in colonoscopy. For example, Tian \etal \cite{tian2022contrastive} recognise abnormal frames from colonoscopy videos from an out-of-distribution view. DAFON \cite{luo2024dynamic} solve an open-set classification problem within a few-shot framework.

\subsection{Detection models}\label{sec:detection_models}


\noindent$\bullet$~\textbf{Input phase.} Detection models classify targets and locate them using boxes, assisting surgical intervention and planning.
The goals of interest are diverse, such as identifying the polyp(s) in images \cite{yang2020colon,liu2021consolidated,qadir2021toward,pacal2022efficient,liu2022source,gong2023frcnn,haugland2023deep,li2023scan++,liu2024decoupled}/videos \cite{tajbakhsh2015comprehensive,tajbakhsh2015automated,yu2016integrating,mo2018efficient,qadir2019improving,zhang2020asynchronous,wu2021multi,yu2022end,jiang2023yona,wang2022explainable,intrator2023self,jiang2024let}, or locating multiple findings \cite{poon2020ai} like bleeding, polyps, and accessory tools.

\noindent$\bullet$~\textbf{Processing phase.} This has three key configurations for the analysis.
\textit{(a) Backbone:} There are two general strategies for network initialisation. The first group \cite{yang2020colon,liu2021consolidated,qadir2021toward,liu2022source,gong2023frcnn,li2023scan++,liu2024decoupled,poon2020ai,wu2021multi,jiang2023yona,intrator2023self,jiang2024let} leverages the ResNet series \cite{he2016deep} pre-trained on the ImageNet \cite{deng2009imagenet} dataset. The second group relies on well-trained object detectors, such as \cite{pacal2022efficient,zhang2020asynchronous} using DarkNet series \cite{redmon2018yolov3,bochkovskiy2020yolov4} and \cite{haugland2023deep} employing EfficientDet-D0 \cite{tan2020efficientdet}.
\textit{(b) Workflow:} Detection models are often built on general-purpose architectures. In the ``WF'' column of \tabref{tab:detection_models}, we categorise the models according to their processing stages. The two-stage workflow decouples detection into the region proposal and classification phases, like models \cite{liu2021consolidated,gong2023frcnn, mo2018efficient,qadir2019improving} based on Faster R-CNN~\cite{ren2015faster}. One-stage models prioritise speed and simplicity, operating in a single forward. For example, some studies \cite{pacal2022efficient,zhang2020asynchronous,poon2020ai} are based on the YOLO framework \cite{redmon2018yolov3,bochkovskiy2020yolov4}, and Yu~\etal~\cite{yu2022end} uses the SSD framework \cite{liu2016ssd}.
\textit{(c) Architecture:} Detection models predict target categories and spatial coordinates, typically implemented in branched frameworks (BF\#1/BF\#2) as shown in \figref{fig:architecture_taxonomy}. Two special cases \cite{qadir2021toward,yu2016integrating} adapt the design of multistream framework to first pop out pixel-wise attention regions, then convert them to bounding boxes.

\begin{table*}[t!]
\centering
\fontsize{6.2pt}{7pt}\selectfont
\renewcommand{\arraystretch}{0.8}
\renewcommand{\tabcolsep}{0.28mm}
\caption{\textbf{Summary of segmentation models in colonoscopy.} 
\underline{Dataset:}
C6 \cite{bernal2015wm}, 
ES \cite{vazquez2017benchmark}, 
KS=Kvasir-SEG \cite{jha2020kvasir}, 
C3 \cite{bernal2012towards}, 
ETIS \cite{silva2014toward}, 
HK \cite{Borgli2020}, 
ASU \cite{tajbakhsh2015automated}, 
CDB \cite{angermann2017towards}, 
BKAI=BKAI-Small \cite{ngoc2021neounet},  
KSe \cite{jha2021comprehensive}, 
GI=GIANA \cite{bernal2017comparative}, 
SUN-S \cite{ji2022video}, 
PG=PolypGen \cite{ali2021polypgen}, K-I=Kvasir-Instrument \cite{jha2021kvasir}.
%
\underline{Backbone:} 
ResUNet \cite{zhang2018road}, 
R-18/-34/-50/-101 \cite{he2016deep}, 
R-50v2 \cite{he2016identity}, 
R2-50=Res2Net-50 \cite{gao2019res2net}, 
V-16 \cite{simonyan2014very}, 
DeiT \cite{touvron2021training},
Eff-B4=EfficientNet -B4 \cite{tan2019efficientnet},
DLV3+=DeepLab V3+ \cite{chen2018encoder},
PB2/3/5=PVTv2-B2/-B3/-B5 \cite{wang2022pvtv2}, 
CvT \cite{wu2021cvt}, 
MiT-B2 \cite{xie2021segformer}, 
CMLP=CycleMLP \cite{chen2021cyclemlp}, 
P-DETR= Point DETR \cite{chen2021points},
D-121=DenseNet121 \cite{huang2017densely},
MN= MSCAN \cite{guo2022segnext},
Swin-T \cite{liu2021swin},
SAM \cite{kirillov2023segment}, SAM2 \cite{ravi2024sam2},
ViT-B16 \cite{dosovitskiy2021vit}, 
DLV2= DeepLabV2 \cite{chen2017deeplab},
HR-W48=HRNet-W48 \cite{wang2020deep},
CN-T=ConvNeXt-Tiny \cite{liu2022convnet},
SFB3=SegFormer-B3 \cite{xie2021segformer}, M2Former=Mask2Former \cite{cheng2022masked}.
%
%
\underline{Edge}-sensitive analysis by explicitly (EX) using edge map as supervision or input and implicitly exploring edge-aware representation (IM\#1) or edge-aware uncertainty (IM\#2).
%
}
\label{tab:segmentation_models}
\begin{tabular}{c | r | r | c | cc | ccc | c | c}
    & Model &Publication &Core design &Training dataset &Testing dataset & Backbone &Arch & Edge &Sup  &URL \\
\hline
    \multirow{80}{*}{\begin{sideways}Image-based models\end{sideways}}
    & Yuan~\etal~\cite{yuan2017automatic} & JBHI'17 & weak bottom-up \& strong top-down saliency & Private & C6 & Customised & BF\#1 & - & US & - \\
    & SFA~\cite{fang2019selective} & MICCAI'19 & area \& boundary constraints & ES & ES & Customised & BF\#1 & EX & FS & \href{https://github.com/Yuqi-cuhk/Polyp-Seg}{Link} \\
    & ResUNet++~\cite{jha2019resunet++} & ISM'19 & enhanced deep residual UNet \cite{zhang2018road} & C6, KS & C6, KS & ResUNet & MF\#1 & - & FS & \href{https://github.com/DebeshJha/ResUNetplusplus}{Link} \\
    & ACSNet~\cite{zhang2020adaptive} & MICCAI'20 & adaptive context selection & ES, KS & ES, KS & R-34 & BF\#1 & - & FS & \href{https://github.com/ReaFly/ACSNet}{Link}\\
    & PraNet~\cite{fan2020pranet} & MICCAI'20 & reverse attention, parallel partial decoder & C6, KS & C6, ES, KS, C3, ETIS & R2-50 & BF\#1 & EX, IM\#1 & FS & \href{https://github.com/DengPingFan/PraNet}{Link} \\ 
    & UI-CNN \cite{wickstrom2020uncertainty} & MedIA'20 & Monte Carlo guided back-propagation & ES & ES & V-16 & MF\#1 & IM\#2 & FS & \href{https://github.com/Wickstrom/uc-in-xai}{Link} \\
    & ThresholdNet~\cite{guo2020learn} & TMI'20 & confidence-guided manifold mixup & ES & ES, WCE-Polyp \cite{guo2020learn} & R-101 & BF\#2 & EX, IM\#1 & FS & \href{https://github.com/Guo-Xiaoqing/ThresholdNet}{Link}\\ 
    & SCR-Net~\cite{wu2021precise} & AAAI'21 & semantic calibration \& refinement & KS & KS & Customised & MF\#1 & - & FS & \href{https://github.com/jiafuz/SCR-Net}{Link} \\
    & BI-GCN~\cite{meng2021bi} & BMVC'21 & boundary-aware graph convolution & C6, KS & C6, ES, KS, C3, ETIS & R2-50 & BF\#1 & EX & FS & \href{https://github.com/smallmax00/BI-GConv}{Link} \\ 
    & FDSemi \cite{wu2021collaborative} & ICCV'21 & collaborative \& adversarial learning & C6, KS & C6, KS & Customised & BF\#2 & IM\#1 & SS & \href{https://github.com/C-GLGLGL/FDSemi}{Link} \\
    & CCBANet~\cite{nguyen2021ccbanet} & MICCAI'21 & cascading context \& balancing attention  & C6, ES, KS & C6, ES, KS & R-34 & BF\#1 & IM\#1 & FS & \href{https://github.com/ntcongvn/CCBANet}{Link} \\
    & CCD~\cite{tian2021constrained} & MICCAI'21 & constrained contrastive distribution learning & HK, Liu \etal \cite{liu2020photoshopping} & HK, Liu \etal \cite{liu2020photoshopping} & R-18 & BF\#2 & - & US & \href{https://github.com/tianyu0207/CCD}{Link} \\
    & HRENet~\cite{shen2021hrenet} & MICCAI'21 & hard region enhancement & C6, KS & C6, KS, C3 & R-34 & BF\#1 & EX & FS & \href{https://github.com/CathySH/HRENet}{Link} \\
    & LOD-Net~\cite{cheng2021learnable} & MICCAI'21 & learnable oriented derivatives & C6, KS & C6, ES, KS, C3, ETIS & R-50 & MF\#2 & EX & FS & \href{https://github.com/midsdsy/LOD-Net}{Link} \\
    & MSNet~\cite{zhao2021automatic} & MICCAI'21 & multiscale subtraction network & C6, KS & C6, ES, KS, C3, ETIS & R2-50 & MF\#1 & EX, IM\#1 & FS & \href{https://github.com/Xiaoqi-Zhao-DLUT/MSNet-M2SNet}{Link}\\
    & SANet~\cite{wei2021shallow} & MICCAI'21 & colour exchange \& probability correction & C6, KS & C6, ES, KS, C3, ETIS & R2-50 & MF\#2 & - & FS & \href{https://github.com/weijun88/SANet}{Link}\\
    & Transfuse~\cite{zhang2021transfuse} & MICCAI'21 & fusing transformers and CNNs & C6, KS & C6, ES, KS, C3, ETIS & R-50v2, ViT-B16 & BF\#2 & EX & FS & \href{https://github.com/Rayicer/TransFuse}{Link} \\
    & EndoUDA \cite{celik2021endouda} & MICCAI'21 & domain adaptation, variational autoencoder training & Celik \etal \cite{celik2021endouda} & Celik \etal \cite{celik2021endouda} & Eff-B4 & BF\#2 & EX & FS, US & \href{https://github.com/sharib-vision/EndoUDA}{Link}\\
    & UACANet~\cite{kim2021uacanet} & MM'21 & uncertainty augmented context attention & C6, KS & C6, ES, KS, C3, ETIS & R2-50 & BF\#1 & EX, IM\#2 & FS & \href{https://github.com/plemeri/UACANet}{Link} \\
    & Jha~\etal~\cite{jha2021comprehensive} & JBHI'21 & \makecell{ResUNet++~\cite{jha2019resunet++} with conditional \\random field \& test-time augmentation} & \makecell{C6, C3, ETIS, \\KS, ASU, CDB} & \makecell{C6, C3, ETIS,\\KS, ASU, CDB} & ResUNet & MF\#1 & - & FS & \href{https://github.com/DebeshJha/ResUNetPlusPlus-with-CRF-and-TTA}{Link} \\
    & MPA-DA~\cite{yang2021mutual} & JBHI'21 & mutual-prototype adaptation network & ES, KS & ETIS & R-101 & BF\#2 & IM\#2 & FS, US & \href{https://github.com/CityU-AIM-Group/MPA-DA}{Link} \\
    & DW-HieraSeg~\cite{guo2021dynamic} & MedIA'21 & hierarchical segmentation, dynamic-weighting & ES & ES & DLV3+ & BF\#1 & - & FS & \href{https://github.com/CityU-AIM-Group/DW-HieraSeg}{Link} \\
    & ICGNet~\cite{du2022icgnet} & IJCAI'22 & context-based reverse-contour guidance & ES, KS & ES, C3, KS & R-34 & BF\#1 & EX & FS & - \\
    & BoxPolyp~\cite{wei2022boxpolyp} & MICCAI'22 & segmentation with extra box labels & C6, KS & C6, ES, KS, C3, ETIS & R2-50, PB2 & BF\#1 & - & WS & \href{https://github.com/weijun88/BoxPolyp}{Link} \\
    & LDNet~\cite{zhang2022lesion} & MICCAI'22 & dynamic kernel generation \& update & Private, C6, KS & Private, C6, KS, C3, ETIS & R2-50 & BF\#1 & - & FS & \href{https://github.com/ReaFly/LDNet}{Link}\\
    & PPFormer~\cite{cai2022using} & MICCAI'22 & polyp-guided self-attention, local-to-global method & C6, KS & C6, ES, KS, C3, ETIS & V-16, CvT & BF\#1 & EX & FS & - \\
    & SSFormer~\cite{wang2022stepwise} & MICCAI'22 & stepwise local \& global feature aggregation & C6, KS & C6, C3, ETIS, KS & MiT-B2 & MF\#1 & - & FS & \href{https://github.com/Qiming-Huang/ssformer}{Link} \\
%
%
    & TRFR-Net~\cite{shen2022task} & MICCAI'22 & task-relevant feature replenishment & C3, ETIS, KS & C3, ETIS, KS & R-34 & BF\#2 & - & FS, US & \href{https://github.com/CathyS1996/TRFRNet}{Link} \\
    & AFP-Mask~\cite{wang2022afp} & JBHI'22 & anchor-free instance segmentation & Private, GI & Private, C6, ETIS & Customised & BF\#1 & - & FS & - \\ 
    & BCNet~\cite{yue2022boundary} & JBHI'22 & cross-layer integration, bilateral boundary extraction & KS & C6, ES, KS & R2-50 & BF\#1 & EX & FS & - \\
    & BSCA-Net~\cite{lin2022bsca} & PR'22 & bit slice context attention & C6, KS & C6, ES, KS, C3, ETIS & R2-50 & BF\#1 & EX & FS & \href{https://github.com/guobaoxiao/BSCANet}{Link} \\
    & Polyp-Mixer~\cite{shi2022polyp} & TCSVT'22 & context-aware MLP-based network & C6, KS & C6, KS, C3, ETIS & CMLP & BF\#1 & - & FS & \href{https://github.com/shijinghuihub/Polyp-Mixer}{Link} \\
%
%
    & ACL-Net~\cite{wu2023acl} & AAAI'23 & affinity contrastive learning & C6, KS & C6, ES, KS, C3, ETIS & R-50 & BF\#2 & - & SS & \href{https://github.com/xiewende/ACL-Net}{Link} \\
    & WS-DefSegNet \cite{ren2023towards} & CVPRw'23 & deformable transformer, sparse foreground loss & W-Polyp \cite{ren2023towards} & C6, KS, C3, ETIS, ES & R2-50 & BF\#1 & - & WS, SS & \href{https://github.com/ic-qialanqian/WS-DefSegNet}{Link} \\
    & WeakPolyp~\cite{wei2023weakpolyp} & MICCAI'23 & mask-to-box transformation, scale consistency & SUN-S, Private & SUN-S, Private & PB2 & BF\#2 & - & WS & \href{https://github.com/weijun88/WeakPolyp}{Link} \\ 
    & PETNet~\cite{ling2023probabilistic} & MICCAI'23 & Gaussian-probabilistic guided semantic fusion & C6, KS & C6, ES, KS, C3, ETIS & PB2 & BF\#1 & - & FS & \href{https://github.com/Seasonsling/PETNet}{Link} \\ 
    & S$^2$ME~\cite{wang2023s} & MICCAI'23 & spatial-spectral mutual teaching, ensemble learning & SUN-S & C6, KS, SUN-S, PG & Customised & BF\#1 & - & WS & \href{https://github.com/lofrienger/S2ME}{Link} \\
    & Su \etal~\cite{su2023revisiting} & MICCAI'23 & feature propagation \& aggregation & C6, KS & C6, ES, KS, C3, ETIS & PB2 & BF\#1 & EX & FS & - \\ 
    & Polyp-PVT~\cite{dong2023polyp} & AIR'23 & Improved pyramid vision transformer & C6, KS & C6, ES, KS, C3, ETIS & PB2 & BF\#1 & EX & FS & \href{https://github.com/DengPingFan/Polyp-PVT}{Link} \\
    & RPANet~\cite{wang2023unsupervised} & IPMI'23 & coarse-to-fine self-supervision & Private & C6, ETIS, KS & R-101 & BF\#2 & - & FS, US & - \\
    & FEGNet \cite{jin2023fegnet} & JBHI'23 & feedback enhancement gate network & C6, KS & C6, ES, KS, C3, ETIS & R2-50 & BF\#1 & EX & FS & - \\ 
    & BS-Loss \cite{du2023boundary} & JBHI'23 & boundary-sensitive loss with location constraint & K-I & K-I & Customised & MF\#1 & EX & FS & \href{https://github.com/dujie-szu/BS-Loss}{Link} \\
    & Point SEGTR~\cite{shi2023deep} & MedIA'23 & multi-point and symmetric consistency & C6, ETIS & C6, ETIS & P-DETR & BF\#2 & - & FS, WS, SS & - \\
    & DGNet~\cite{ji2023deep} & MIR'23  & deep gradient learning & C6, KS & C3, ETIS & Eff-B4 & BF\#1 & EX & FS & \href{https://github.com/GewelsJI/DGNet}{Link}\\
    & CFA-Net \cite{zhou2023cross} & PR'23 & cross-level feature fusion, boundary aggregation & C6, KS & C6, ES, KS, C3, ETIS & R2-50 & BF\#1 & EX & FS & \href{https://github.com/taozh2017/CFANet}{Link} \\ 
    & ColnNet \cite{jain2023coinnet} & TMI'23 & statistical attention, anomaly boundary approximation & C6, KS & C6, ES, KS, C3, ETIS & D-121 & BF\#1 & EX & FS & - \\
    & FANet \cite{tomar2023fanet} & TNNLS'23 & feedback attention, run-length encoding & C6, KS & C6, KS & Customised & MF\#2 & - & FS & \href{https://github.com/nikhilroxtomar/FANet}{Link} \\
%
%
%
    & MCANet~\cite{shao2023mcanet} & arXiv'23 & multi-scale cross-axis attention & C6, KS & C6, ES, KS, C3, ETIS & MN & MF\#1 & - & FS & \href{https://github.com/haoshao-nku/medical_seg/tree/master/mmsegmentation/local_config/MCANet}{Link}\\
    & Polyper~\cite{shao2024polyper} & AAAI'24 & boundary sensitive attention & C6, KS & C6, KS, C3, ETIS, ES & Swin-T & MF\#1 & IM\#1 & FS & \href{https://github.com/haoshao-nku/medical_seg/tree/master/mmsegmentation/local_config/Polyper-AAAI2024}{Link} \\
    & EMCAD~\cite{rahman2024emcad} & CVPR'24 & efficient multi-scale convolutional attention decoding & C6, C3, ETIS, KS, BKAI & C6, C3, ETIS, KS, BKAI & PB2 & BF\#1 & - & FS & \href{https://github.com/SLDGroup/EMCAD}{Link} \\
    & Sch{\"o}n \etal \cite{schon2024adapting} & CVPR'24 & SAM \cite{kirillov2023segment}, test-time adaptation & K-I, CDB, KS & K-I, CDB, KS & SAM & BF\#2 & EX & WS & - \\
%
    & MH-pFLID~\cite{xie2024mh} & ICML'24 & federated learning, injection-distillation paradigm & Private & Private & Customised & BF\#2 & - & FS & \href{https://github.com/xiely-123/MH-FLIDv1}{Link} \\
    & ASPS \cite{li2024asps} & MICCAI'24 & SAM \cite{kirillov2023segment}, uncertainty-guided regularisation & C6, KS & C6, ES, KS, C3, ETIS & SAM, MN & BF\#2 & IM\#2 & FS & \href{https://github.com/HuiqianLi/ASPS}{Link} \\
    & Polyp-Mamba~\cite{Xu_PolypMamba_MICCAI2024} & MICCAI'24 & vision state space model, semantic relationship mining  & C6, KS & C6, ES, KS, C3, ETIS & VMamba~\cite{liu2024vmamba} & MF\#1 & EX & FS & - \\
    & QueryNet~\cite{Cha_QueryNet_MICCAI2024} & MICCAI'24 & unified framework of segmentation \& detection & C6, KS & C6, ES, KS, C3, ETIS & M2Former & MF\#2 & - & FS & \href{https://github.com/JiaxingChai/Query_Net}{Link} \\
%
%
    & LSSNet \cite{Wan_LSSNet_MICCAI2024} & MICCAI'24 &  local \& shallow feature supplementation & C6, KS & C6, ES, KS, C3, ETIS & PB2 & BF\#1 & EX & FS & \href{https://github.com/heyeying/LSSNet}{Link} \\
    & BSBP-RWKV~\cite{zhou2024bsbp} & MM'24 & Perona-Malik diffusion, RWKV \cite{peng2023rwkv} & KS & KS & Customised & BF\#1 & EX & FS & -\\
    & CFATransUnet~\cite{wang2024cfatransunet} & CIBM'24 & channel-wise cross fusion attention and transformer & C6, KS & C6, KS & PB3 & BF\#1 & - & FS & \href{https://github.com/CPU0808066/CFATransUnet}{Link} \\
    & PolypMixNet~\cite{jia2024polypmixnet} & CIBM'24 & consistency regularisation, soft pseudo labeling & C3, C6, KS, ETIS  & C3, C6, KS, ETIS  & R-34 & BF\#2 & - & SS & \href{https://github.com/YChienHung/PolypMix}{Link} \\
    & RGIAug~\cite{zhang2024generalizable} & JBHI'24 & randomised global illumination augmentation & C3, C6, ETIS, KS & C3, C6, ETIS, KS & Customised & BF\#2 & - & FS & \href{https://github.com/ChangJoey56/RGIAug}{Link} \\
    & EMTS-Net~\cite{wang2024efficient} & JBHI'24 & multi-task synergetic network & C6, KS & C6, ES, KS, C3, ETIS & R2-50 & BF\#1 & - & FS & - \\
    & MSDE-Net \cite{yang2024msde} & JBHI'24 & multi-scale dual-encoding network & K-I & K-I & R-34, DeiT & MF\#2 & - & FS & -\\
    & Polyp-OOD~\cite{ji2023rethinking} & MIR'24 & out-of-distribution modelling, latent standardisation & SUN-S & SUN-S, C6, C3, ETIS, KS & ViT-B16 & MF\#2 & - & US & \href{https://github.com/GewelsJI/Polyp-OOD}{Link} \\
    & MedSAM~\cite{ma2024segment} & NComs'24 & SAM \cite{kirillov2023segment}, cross-organ/modality tuning & Hybrid datasets & Hybrid datasets & SAM & MF\#2 & - & WS & \href{https://github.com/bowang-lab/MedSAM}{Link} \\
    & FoBS~\cite{liu2024devil} & TCSVT'24 & multi-level boundary-enhanced framework & KS, ES & KS, ES, ETIS, C3 & DLV3+ & BF\#2 & EX, IM\#1 & FS & \href{https://github.com/TFboys-lzz/FoBS}{Link}\\
    & DCL-PS~\cite{lu2024domain} & TMI'24 & domain-interactive contrastive learning, self-training & ETIS, HK, ES, KS & ES, KS & DLV2 & BF\#2 & - & FS, US & \href{https://github.com/taozh2017/DCLPS}{Link} \\
%
    & Gao \etal \cite{gao2024boosting} & TMI'24 & in-context learning, dual similarity checkup & C3 & C3 & SAM & BF\#2 &- & WS & \\
%
    & SliceMamba \cite{fan2024slicemamba} & arXiv'24 & vision state space model, bidirectional slice scan & C6, KS & C6, KS & Customised & MF\#1 & - & FS & - \\
    & ProMamba~\cite{xie2024promamba} & arXiv'24 & vision state space model, promptable segmentation & C6, KS & C6, KS, C3, ETIS, ES, BKAI & Vim \cite{zhu2024vision}& MF\#2 & - & WS & - \\
    & SAM2-UNet \cite{xiong2024sam2} & arXiv'24 & SAM2 \cite{ravi2024sam2}, adapter-based tuning  & C6, KS & C6, ES, KS, C3, ETIS  & SAM2 & BF\#1 & EX & FS & \href{https://github.com/WZH0120/SAM2-UNet}{Link} \\
    & U-KAN~\cite{li2025u} & AAAI'25 & U-shaped Kolmogorov-Arnold network \cite{liu2024kan} & C6 & C6 & Customised & MF\#1 & - & FS & \href{https://yes-u-kan.github.io/}{Link} \\
%
%
%
\hline
    \multirow{16}{*}{\begin{sideways}Video-based models\end{sideways}}
    & Puyal~\etal~\cite{puyal2020endoscopic} & MICCAI'20 & temporal correlation via hybrid 2D/3D CNNs  & Private, KS & Private & R-101 & MF\#2 & - & FS & - \\
    & PNS-Net~\cite{ji2021progressively} & MICCAI'21 & normalised self-attention, progressive learning & C6, C3, ASU, KS & C6, C3 & R2-50 & BF\#2 & - & FS & \href{https://github.com/GewelsJI/PNS-Net}{Link} \\
    & SSTAN ~\cite{zhao2022semi} & MICCAI'22 & spatial-temporal attention & LDPolyVideo \cite{ma2021ldpolypvideo} & LDPolyVideo \cite{ma2021ldpolypvideo} & ResUNet & BF\#2 & - & SS & \href{https://github.com/ShinkaiZ/SSTAN}{Link} \\
    & TCCNet~\cite{li2022tccnet} & IJCAI'22 & temporal consistency, context-free loss & C6, C3 & C6, C3, ETIS & R2-50 & BF\#2 & EX, IM\#1 & SS & \href{https://github.com/wener-yung/TCCNet}{Link} \\
    & Puyal~\etal~\cite{puyal2022polyp} & MedIA'22 & extend \cite{puyal2020endoscopic} with optimal setups & Private, KS & Private, SUN & R-101 & BF\#2 & - & FS & - \\
    & PNS+~\cite{ji2022video} & MIR'22 & extend \cite{ji2021progressively} with global-to-local learning & SUN-S & SUN-S & R2-50 & BF\#2 & - & FS & \href{https://github.com/GewelsJI/VPS}{Link} \\
    & EUVPS~\cite{fang2024embedding} & AAAI'24 & cross-scale region linking, cross-wise scale alignment & SUN-S, C6 & SUN-S, C6 & HR-W48 & BF\#2 & - & FS & \href{https://github.com/zhixue-fang/EUVPS}{Link} \\
    & LGRNet \cite{xu2024lgrnet} & MICCAI'24 & cyclic neighbourhood propagate, Hilbert selective scan & C6, C3, SUN-S & C6, C3, SUN-S &  R2-50 & BF\#2 & - & FS & \href{https://github.com/bio-mlhui/LGRNet}{Link}\\
    & SALI \cite{hu2024sali} & MICCAI'24 & short-term alignment, long-term interaction module & SUN-S & SUN-S & PB5 & BF\#2 & EX & FS & \href{https://github.com/Scatteredrain/SALI}{Link}\\
    & Diff-VPS \cite{lu2024diff} & MICCAI'24 & diffusion model, adversarial temporal reasoning & SUN-S & SUN-S & SFB3 & BF\#2 & - & FS, US & \href{https://github.com/lydia-yllu/Diff-VPS}{Link} \\
    & FlowICBNet~\cite{wan2024iterative} & CIBM'24 & iterative feedback units, frame filtering \& selecting & SUN-S & SUN-S & R2-50 & BF\#2 & - & FS & \href{https://github.com/eraserNut/ICBNet}{Link} \\
%
%
    & Drag\&Drop~\cite{chou2024acquiring} & MIR'24 & weakly-supervised temporal annotator & SUN-S & SUN-S & - & BF\#2 & - & WS & \href{https://github.com/johnson111788/Drag-Drop}{Link} \\
    & SSTFB \cite{xu2024sstfb} & arXiv'24 & self-supervised encoder, sub-branching mechanism & SUN-S & SUN-S, CDB & R2-50 & BF\#2 & - & US, FS  & - \\
    & Vivim~\cite{yang2024vivim} & arXiv'24 & video state space model, spatio-temporal selective scan & KS, ASU, C3, C6 & C3, C6 & Customised & BF\#2 & EX & FS & \href{https://github.com/scott-yjyang/Vivim}{Link}\\
    & MAST \cite{chen2024mast} & arXiv'24 & Siamese transformer, mixture attention module & SUN-S & SUN-S & PB2 & BF\#2 & - & FS & \href{https://github.com/Junqing-Yang/MAST}{Link} \\
%
%
%
\end{tabular}
\end{table*}

\noindent$\bullet$~\textbf{Output phase.}
{\textit{(a) Post-processing techniques} are employed to eliminate duplicate predictions and select the most relevant targets, with non-maximum suppression (NMS) being a widely-adopted method \cite{pacal2022efficient,liu2022source,gong2023frcnn,haugland2023deep,mo2018efficient,qadir2019improving,poon2020ai,yu2022end,jiang2023yona}.}
\textit{(b) Auxiliary information} can improve prediction reliability, such as edge cues suggested to provide geometric patterns for object detection in \cite{qadir2021toward,tajbakhsh2015comprehensive,tajbakhsh2015automated}.
\textit{(c) Supervision strategy} is currently dominated by fully supervised learning, such as using region-level labels \cite{haugland2023deep,pacal2022efficient,yang2020colon,gong2023frcnn,mo2018efficient,qadir2019improving,jiang2024let,wu2021multi,wang2022explainable,zhang2020asynchronous,yu2022end} and pixel-wise \cite{yu2016integrating,qadir2021toward,tajbakhsh2015comprehensive,tajbakhsh2015automated} labels, and introducing the box-assisted contrastive learning \cite{jiang2023yona}. Other models \cite{liu2021consolidated,liu2022source,li2023scan++,liu2024decoupled} explore unsupervised domain adaptation techniques to detect polyps across colonoscopy devices.

\noindent$\bullet$~\textbf{Remarks.} We emphasise a few observations for the above review. 
\textit{(a)} Most models focus on detecting polyp(s), while other colonoscopic findings receive less attention. We encourage exploring public multitarget \cite{hoang2019enhancing} or multicentre \cite{jha2024polypdb} data.
\textit{(b)} Learning strategies are underexplored. General-purpose detection models 
like using weak supervision \cite{zhou2016learning} provide valuable references, being potentially more feasible and cost-effective since they require less detailed annotations from medical experts.
\textit{(c)} Beyond well-established convolution-based detection frameworks, recent methods like transformer-based DETR \cite{carion2020end} and diffusion-based DiffusionDet \cite{chen2023diffusiondet}, open exciting opportunities for this field. Moreover, exploring cross-task synergy is promising, as three video-based models \cite{zhang2020asynchronous,yu2022end,intrator2023self} demonstrated effectiveness in unifying polyp detection and tracking frameworks.
\textit{(d)} Although some datasets such as the SUN-database \cite{misawa2021development} ($>$158K samples) and LDPolyVideo \cite{ma2021ldpolypvideo} ($>$900K samples) are relatively large, this field still lacks a standardised evaluation benchmark.

\begin{table*}[t!]
\centering
\scriptsize
\renewcommand{\arraystretch}{1.05}
\renewcommand{\tabcolsep}{0.6mm}
\caption{\textbf{Comparison of image polyp segmentation models.} Models are evaluated using the mean scores (\%) of structure measure ($\mathcal{S}$ \cite{cheng2021smeasure}) and Dice coefficient ($\mathcal{D}$) on two test sets, with boxplots illustrating the distribution of their consistency and variability across test cases. The rankings for each model are marked in \tgray{grey} colour.
}
\label{tab:seg_model_ranking}
\begin{tabular}{r|l|c|l|c|l|c|l|c}
& \multicolumn{4}{c|}{Kvasir-SEG (100 test images) \cite{jha2020kvasir}} & \multicolumn{4}{c}{CVC-ClinicDB (62 test images) \cite{bernal2015wm}} \\
\cline{2-9}
Model 
& $\mathcal{S}$ \tiny{\tgray{[\#Rank]}} 
& 0.0~~~$\rightharpoonup$~~~~0.5~~~~$\rightharpoonup$~~~1.0 
& $\mathcal{D}$ \tiny{\tgray{[\#Rank]}} 
& 0.0~~~$\rightharpoonup$~~~~0.5~~~~$\rightharpoonup$~~~1.0 
& $\mathcal{S}$ \tiny{\tgray{[\#Rank]}} 
& 0.0~~~$\rightharpoonup$~~~~0.5~~~~$\rightharpoonup$~~~1.0 
& $\mathcal{D}$ \tiny{\tgray{[\#Rank]}} 
& 0.0~~~$\rightharpoonup$~~~~0.5~~~~$\rightharpoonup$~~~1.0 \\
\hline
Polyp-PVT~\cite{dong2023polyp} 
& 92.51~\tiny{\tgray{[\#1]}} 
& \multirow{14}{*}{\includegraphics[height=4.14cm]{Figures/tab5_KS_S-min.pdf}}
& 91.74~\tiny{\tgray{[\#2]}}
& \multirow{14}{*}{\includegraphics[height=4.14cm]{Figures/tab5_KS_dice-min.pdf}}
& 95.00~\tiny{\tgray{[\#2]}}
& \multirow{14}{*}{\includegraphics[height=4.14cm]{Figures/tab5_C6_S-min.pdf}}
& 93.68~\tiny{\tgray{[\#1]}}
&\multirow{14}{*}{\includegraphics[height=4.14cm]{Figures/tab5_C6_dice-min.pdf}} \\
CFA-Net~\cite{zhou2023cross} 
& 92.40~\tiny{\tgray{[\#2]}} & 
& 91.47~\tiny{\tgray{[\#4]}} & 
& 95.07~\tiny{\tgray{[\#1]}} &
& 93.25~\tiny{\tgray{[\#2]}} &\\
MSNet~\cite{zhao2021automatic} 
& 92.31~\tiny{\tgray{[\#3]}} & 
& 90.23~\tiny{\tgray{[\#7]}} & 
& 94.68~\tiny{\tgray{[\#3]}} &
& 91.48~\tiny{\tgray{[\#6]}} &\\
BoxPolyp~\cite{wei2022boxpolyp} 
& 92.30~\tiny{\tgray{[\#4]}} & 
& 91.84~\tiny{\tgray{[\#1]}} & 
& 93.70~\tiny{\tgray{[\#6]}} &
& 91.81~\tiny{\tgray{[\#4]}} &\\
SSFormer~\cite{wang2022stepwise} 
& 92.21~\tiny{\tgray{[\#5]}} & 
& 91.71~\tiny{\tgray{[\#3]}} & 
& 92.87~\tiny{\tgray{[\#9]}} &
& 90.60~\tiny{\tgray{[\#7]}} &\\
UACANet~\cite{kim2021uacanet} 
& 91.67~\tiny{\tgray{[\#6]}} & 
& 91.21~\tiny{\tgray{[\#5]}} & 
& 94.30~\tiny{\tgray{[\#4]}} &
& 92.63~\tiny{\tgray{[\#3]}} &\\
PraNet~\cite{fan2020pranet} 
& 91.50~\tiny{\tgray{[\#7]}} & 
& 89.82~\tiny{\tgray{[\#8]}} & 
& 93.68~\tiny{\tgray{[\#7]}} &
& 89.90~\tiny{\tgray{[\#9]}} &\\
SANet~\cite{wei2021shallow} 
& 91.45~\tiny{\tgray{[\#8]}} & 
& 90.41~\tiny{\tgray{[\#6]}} & 
& 93.98~\tiny{\tgray{[\#5]}} &
& 91.57~\tiny{\tgray{[\#5]}} &\\
DGNet~\cite{ji2023deep} 
& 90.98~\tiny{\tgray{[\#9]}} & 
& 89.72~\tiny{\tgray{[\#9]}} & 
& 93.39~\tiny{\tgray{[\#8]}} &
& 90.44~\tiny{\tgray{[\#8]}} &\\
MCANet~\cite{shao2023mcanet} 
& 90.25~\tiny{\tgray{[\#10]}} & 
& 89.55~\tiny{\tgray{[\#10]}} & 
& 91.79~\tiny{\tgray{[\#10]}} &
& 89.70~\tiny{\tgray{[\#10]}} &\\
Polyper~\cite{shao2024polyper} 
& 90.08~\tiny{\tgray{[\#11]}} & 
& 89.12~\tiny{\tgray{[\#11]}} & 
& 91.29~\tiny{\tgray{[\#11]}} &
& 88.63~\tiny{\tgray{[\#11]}} &\\
UNet++~\cite{zhou2019unet++} 
& 86.21~\tiny{\tgray{[\#12]}} & 
& 82.08~\tiny{\tgray{[\#12]}} & 
& 87.33~\tiny{\tgray{[\#13]}} &
& 79.42~\tiny{\tgray{[\#13]}} &\\
UNet~\cite{ronneberger2015u} 
& 85.76~\tiny{\tgray{[\#13]}} & 
& 81.83~\tiny{\tgray{[\#13]}} & 
& 89.00~\tiny{\tgray{[\#12]}} &  
& 82.25~\tiny{\tgray{[\#12]}} &\\
SFA~\cite{fang2019selective} 
& 78.14~\tiny{\tgray{[\#14]}} & 
& 72.31~\tiny{\tgray{[\#14]}} & 
& 79.33~\tiny{\tgray{[\#14]}} &
& 70.06~\tiny{\tgray{[\#14]}} &\\
%
\end{tabular}
\end{table*}

\subsection{Segmentation models}\label{sec:segmentation_models}

Compared to the above two topics, the segmentation research appears to be well-established (see \tabref{tab:segmentation_models}). 

\noindent$\bullet$~\textbf{Input phase.} Most segmentation models focus on a single target (\ie, polyp), typically adopting as binary segmentation paradigm. These models usually follow the well-established testbed of PraNet \cite{fan2020pranet} for their development and comparison. An exception case, AFP-Mask \cite{wang2022afp}, provides an anchor-free framework for segmenting polyp instances. Recent works \cite{du2023boundary,yang2024msde,schon2024adapting} have also focused on segmenting surgical tools during procedures.

\noindent$\bullet$~\textbf{Processing phase.} 
%
\textit{(a) Backbone:} The visual encoders for the segmentation task have been extensively explored. A common option is to use a general backbone pre-trained on ImageNet \cite{deng2009imagenet}, such as using CNN \cite{fan2020pranet,shao2023mcanet}, vision Transformer \cite{dong2023polyp,shao2024polyper}, hybrid CNN-Transformer network \cite{Wan_LSSNet_MICCAI2024,cai2022using,zhang2021transfuse}, multilayer perceptron \cite{shi2022polyp}, state space model \cite{Xu_PolypMamba_MICCAI2024,xie2024promamba,yang2024vivim,fan2024slicemamba}, receptance weighted key value (RWKV) \cite{zhou2024bsbp}, and Kolmogorov-Arnold network \cite{zhou2024bsbp}. 
An alternative is to use well-trained perception models such as Point DETR \cite{chen2021points} used in \cite{shi2023deep}, DeepLab series \cite{chen2018encoder,chen2017deeplab} applied in \cite{guo2021dynamic,liu2024devil,lu2024domain}. Recently, there has been a shift towards promptable architectures. The first way is to exploit the foundation models, for example, by fine-tuning the segment anything modal (SAM) \cite{kirillov2023segment} with location prompts \cite{ma2024segment}, adapting SAM during the test time \cite{schon2024adapting}, exploiting the hybrid CNN-Transformer network \cite{li2024asps}, or incorporating trainable adapter layers into the SAM2's encoder \cite{xiong2024sam2}. Another way aims to adapt the model to unseen scenarios through in-context learning \cite{gao2024boosting}.
\textit{(b) Architecture:}
The community favours the encoder-decoder design for its superior ability to perceive hierarchical features. Current models usually opt for multistream or branched frameworks, as in \figref{fig:architecture_taxonomy}. Various modifications in this area have been explored, such as incorporating residual connected flows \cite{jha2019resunet++,jha2021comprehensive}, probing cross-task synergy \cite{Cha_QueryNet_MICCAI2024}, providing additional edge cues \cite{ji2023deep}, using the model-ensembling strategy \cite{zhang2021transfuse,cai2022using,li2024asps}, calculating latent statistics \cite{ji2023rethinking}, exploring spatio-temporal relationships through 3D convolutions \cite{puyal2020endoscopic,puyal2022polyp} or self-attention modules \cite{ji2021progressively,ji2022video,hu2024sali}, and approaching with the teacher-student paradigm \cite{ren2023towards,shi2023deep}.
\textit{(c) Edge-sensitive analysis:} Geometric patterns are beneficial in enhancing the model's capability to differentiate foreground objects from the background. The current techniques are in two main ways. The first involves the explicit use of edge maps derived from image gradients, either for direct supervision \cite{fang2019selective,shen2021hrenet,ji2023deep} or as an auxiliary input \cite{celik2021endouda}. Moreover, some methods emphasise edge-aware calculation within their loss functions, such as boundary weighted \cite{cai2022using,xiong2024sam2} and customised \cite{cheng2021learnable,du2023boundary}). Second, edge information can be implicitly integrated by embedding edge-aware representations (\eg, reverse attention \cite{fan2020pranet}, morphology operator \cite{shao2024polyper}, subtraction operator \cite{guo2020learn}), or by quantifying edge-aware uncertainties \cite{wickstrom2020uncertainty,kim2021uacanet}. Moreover, some methods adopt a hybrid strategy, \eg, both the subtraction operator and the edge-aware loss are used in MSNet \cite{zhao2021automatic}.

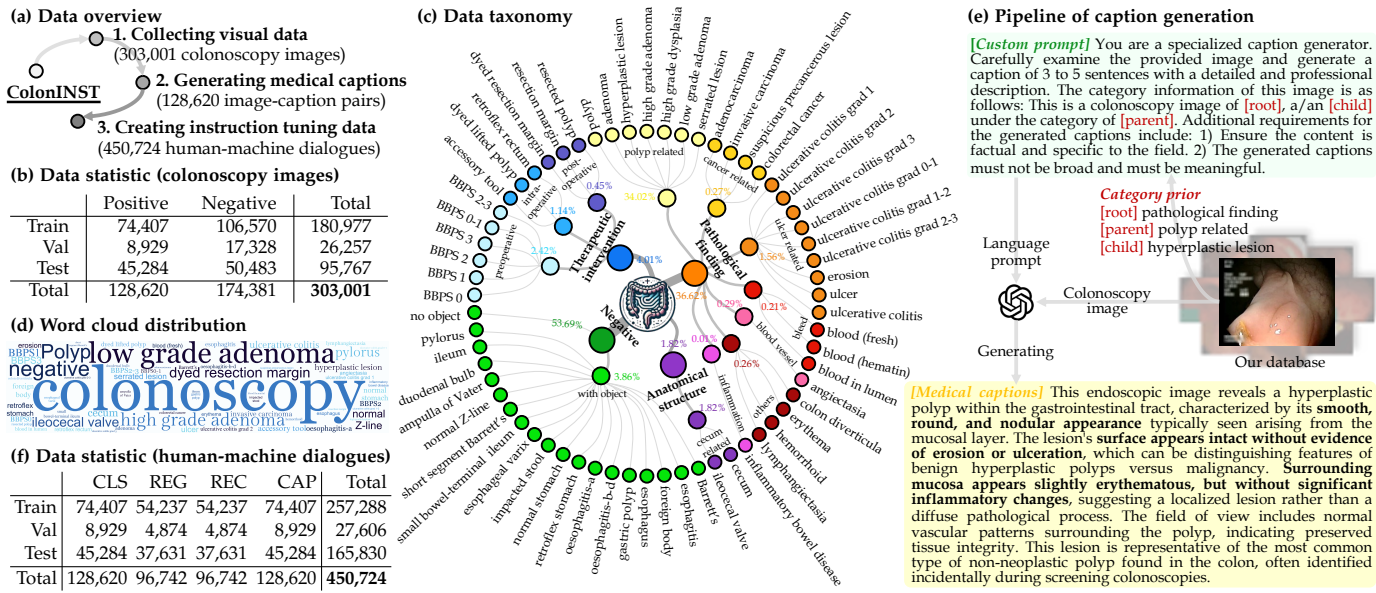
\begin{figure*}[t!]
\centering
\begin{overpic}[width=\textwidth]{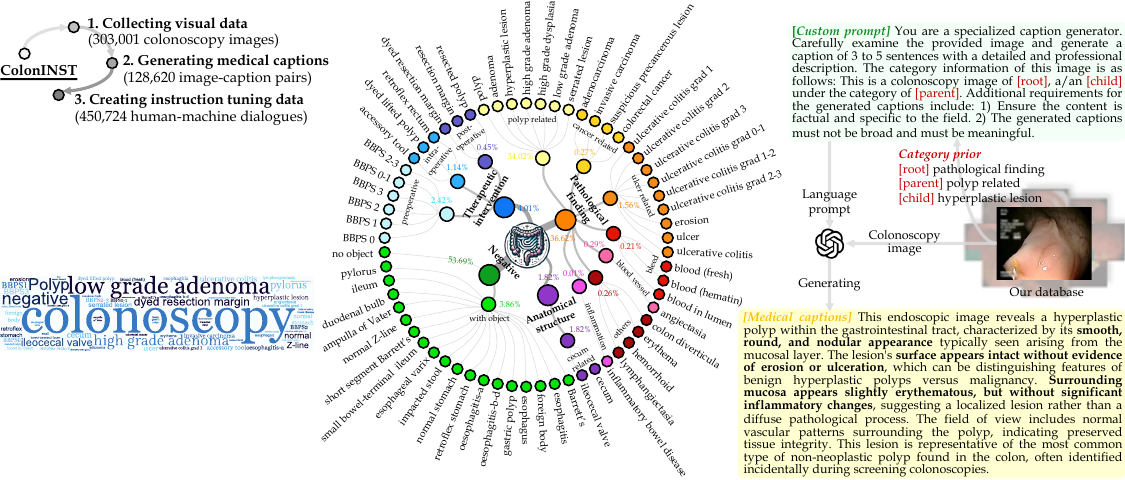}
\put(0.3,41.4){\scriptsize \hypertarget{fig4_a}{\textbf{(a) Data overview}}}
\put(0.3,29.8){\scriptsize \hypertarget{fig4_b}{\textbf{(b) Data statistic (colonoscopy images)}}}
\put(30,41.4){\scriptsize \hypertarget{fig4_c}{\textbf{(c) Data taxonomy}}}
\put(0.3,18.8){\scriptsize \hypertarget{fig4_d}{\textbf{(d) Word cloud distribution}}}
\put(0.3,9.25){\scriptsize \hypertarget{fig4_f}{\textbf{(f) Data statistic (human-machine dialogues)}}}
\put(70.2,41.4){\scriptsize \hypertarget{fig4_e}{\textbf{(e) Pipeline of caption generation}}}
\put(-1,24.5)
{
\scriptsize
\renewcommand{\arraystretch}{1}
\renewcommand{\tabcolsep}{2.2mm}
\begin{tabular}{r|rr|r}
    & Positive  & Negative  & Total \\
    \hline
    Train & 74,407 & 106,570 & 180,977 \\
    Val & 8,929 & 17,328 & 26,257 \\ 
    Test & 45,284 & 50,483 & 95,767 \\ 
    \hline
    Total & 128,620 & 174,381 & \textbf{303,001} \\
\end{tabular}
}
\put(-1,3.85)
{
\scriptsize
\renewcommand{\arraystretch}{1.1}
\renewcommand{\tabcolsep}{0.59mm}
\begin{tabular}{r|rrrr|r}
    & CLS  & REG  & REC & CAP & Total \\
    \hline
    Train & 74,407 & 54,237 & 54,237 & 74,407 & 257,288 \\ 
    Val & 8,929 & 4,874 & 4,874 & 8,929 & 27,606 \\ 
    Test & 45,284 & 37,631 & 37,631 & 45,284 & 165,830 \\ 
    \hline
    Total & 128,620 & 96,742 & 96,742 & 128,620 & \textbf{450,724} \\
\end{tabular}
}
\end{overpic}
\vspace{-15pt}
\caption{\textbf{Details of \ourdata.} 
(a) Three sequential steps to create the instruction tuning dataset for multimodal research. (b) Numbers of colonoscopy images designated for training, validation, and testing purposes. {
(c) Data taxonomy of three-level categories.
(d) A word cloud of the category distribution by name size.  (e) Caption generation pipeline using the VL prompting mode of GPT-4V \cite{openai2023chatgpt4v}. (f) Numbers of human-machine dialogues created for four tasks.}
}
\label{fig:coloninst}
\end{figure*}

\noindent$\bullet$~\textbf{Output phase.} Most models are trained in a fully supervised way, always incorporating deep supervision at various decoding stages, as seen in \cite{fan2020pranet,dong2023polyp}. Recent models have extended beyond with data-efficient approaches, for example, weakly supervised mask-to-box transformation~\cite{wei2023weakpolyp} and unsupervised techniques such as contrastive learning \cite{tian2021constrained}, out-of-distribution modelling \cite{ji2023rethinking}, and pseudo-label supervision \cite{zhao2022semi,wei2022boxpolyp}. Hybrid supervised strategies are also present in which models \cite{celik2021endouda,yang2021mutual,shen2022task,wang2023unsupervised} undergo fully supervised training in the source domain, and are then adapted to the target domain in an unsupervised way. Point SEGTR \cite{shi2023deep} exploits three types of supervision to enhance the model. Additionally, some teacher-student networks receive hybrid supervision signals; for example, Ren \etal \cite{ren2023towards} employs a weakly supervised approach for the teacher while the student undergoes semi-supervised training.

\noindent$\bullet$~\textbf{Remarks.} 
To reflect current field progress, we analysed 14 open-source image segmentation models on two popular test datasets, as shown in \tabref{tab:seg_model_ranking}. All models are trained on the same dataset from Fan \etal's benchmark \cite{fan2020pranet}. First, we observe that current learning strategies are underexplored, as evidenced by BoxPolyp, a weakly supervised model that obtains the best $\mathcal{D}$ score (91.84\%) on Kvasir-SEG. In addition, some models achieve better performance, yet exhibit wider interquartile ranges in boxplots, indicating their variability in predictions. For example, Polyp-PVT, which ranks highest in the $\mathcal{S}$ score on Kvasir-SEG, shows a wider interquartile range than other models like SSFormer. From these results, we suggest several promising opportunities for future study.
\textit{(a)} The current gold benchmark \cite{fan2020pranet} comprises less than 1.5K samples and is focused on a category (polyp). In general, scaling up both the data size and diversity could be a natural way to improve robustness and generalisability. This demand is driving innovations, such as developing a semi-auto image annotator \cite{ravi2024sam2,kirillov2023segment} to reduce expert labour and synthesising high-fidelity content via diffusion \cite{chen2024towards} and autoregressive \cite{tian2024visual} techniques.
\textit{(b)} Moreover, infinite data scaling is not sustainable. Developing data-efficient strategies \cite{zhang2021dodnet,karimi2020deep,yu2014large} that require fewer or weaker labels is more cost-effective for average users in the community.
\textit{(c)} Finally, providing procedural support to physicians is essential, including anomaly detection, navigation planning, risk assessment, and intervention advice. We can adopt innovations from similar topics \cite{hu2024ophnet}.

\subsection{VL models}
\label{sec:vision_language_models}
Compared to the above three topics, multimodal research has relatively fewer references. Most existing methods are discriminative models that aim to learn decision boundaries between multimodal inputs. Some studies demonstrate the model's effectiveness in referring segmentation tasks, for example, by incorporating textual attention of lesion attributes  (\eg, size and number of polyps) into a U-shaped model \cite{tomar2022tganet}, a diffusion model \cite{zhao2024dtan}, or a hybrid network \cite{zhaotact}. Other studies \cite{qin2023medical,guo2023multiple} have developd prompt engineering pipelines based on well-trained GLIP \cite{li2022glip} for polyp detection. 
Moreover, the SAM is capable of operating in a VL setting, obtaining location prompts from either the image-text activation map \cite{wang2023knowledge} or a zero-shot grounding detector \cite{biswas2023polyp,Zha_TextPolyp_MICCAI2024}. In the MEDVQA-GI competition \cite{hicks2023overview}, most solutions are discriminative-based, approaching VL tasks as a classification mapping problem, where predefined labels are assigned to image-text pairs. An alternative is a generative-based solution \cite{wang2023adapting} that adapts a pretrained BLIP-2 model \cite{li2023blip2} to generate predictions.

\noindent$\bullet$~\textbf{Remarks.} 
Two possible reasons might explain the lag in VL research for colonoscopy. \textit{(a) Data-centric issue.} The lack of well-structured and high-quality image-text pairs hinders progress. Future insights can be learnt from existing ideas. First, crawling unlabelled image-text data from social media \cite{huang2023visual} or the scientific literature \cite{taylor2022galactica} can be used to build domain-specific foundation models. Second, language models such as GPT-4V \cite{openai2023chatgpt4v} can generate diverse professional descriptions, offering an economical and scalable way to expand the knowledge space of the data.
\textit{(b) Model-centric issue.} Current VL techniques in colonoscopy have not kept pace, even with recent developments in multimodal language models (MLMs) \cite{liu2024llavav1,liu2024llavav15} for general domains. These techniques opt for a decoder-only strategy that unifies multiple tasks (\eg, detection, captioning) into a unified autoregressive framework (\ie, next-token prediction). These models are flexibly capable of processing input and output texts of varying lengths, without requiring additional task-specific heads for different tasks.

\section{Advancing Multimodal Learning for Colonoscopy}\label{sec:multimodal_solution}

\begin{table*}[t!]
\centering
\fontsize{6.2pt}{7pt}\selectfont
\renewcommand{\arraystretch}{0.6}
\renewcommand{\tabcolsep}{1.35mm}
\caption{\textbf{Details of instruction tuning dataset \ourdata.} For each task, we provide five templates for human instructions, the data sources used to organise human-machine dialogues, and an example of a human-machine conversation.
}
\label{tab:coloninst_dialogue}
\begin{tabular}{c|l|l|l}
Task & Instruction templates & Data source & Human-machine dialogue sample \\
\hline
CLS & \makecell[l]{
    1. Categorize the object. \\
    2. Determine the object's category. \\
    3. Identify the category of the object. \\
    4. Classify the object's category. \\
    5. Assign the object to its corresponding category. \\} 
& \makecell[l]{19 sources $\rightarrow$
    SUN-database \cite{misawa2021development}, PolypGen \cite{ali2021polypgen}, CVC-ClinicDB \cite{bernal2015wm},\\
    ETIS \cite{silva2014toward}, KUMC \cite{li2021colonoscopy}, Kvasir \cite{pogorelov2017kvasir}, PSNBI2K \cite{psnbi2k}, CVC-ColonDB \cite{bernal2012towards},\\
    EDD2020 \cite{ali2020endoscopy}, Kvasir-Capsule \cite{smedsrud2021kvasir}, CP-CHILD \cite{wang2020improved}, BKAI-Small \cite{ngoc2021neounet},\\
    PICCOLO \cite{sanchez2020piccolo}, WCE-CCDD\cite{montalbo2022diagnosing}, CPC-Paired \cite{wang2021colorectal}, HyperKvasir \cite{Borgli2020}, \\
    Nerthus \cite{pogorelov2017nerthus}, GastroVision \cite{jha2023gastrovision}, Kvasi-Instrument \cite{jha2021kvasir}}
& \adjustbox{valign=m}{\includegraphics[height=1.15cm]{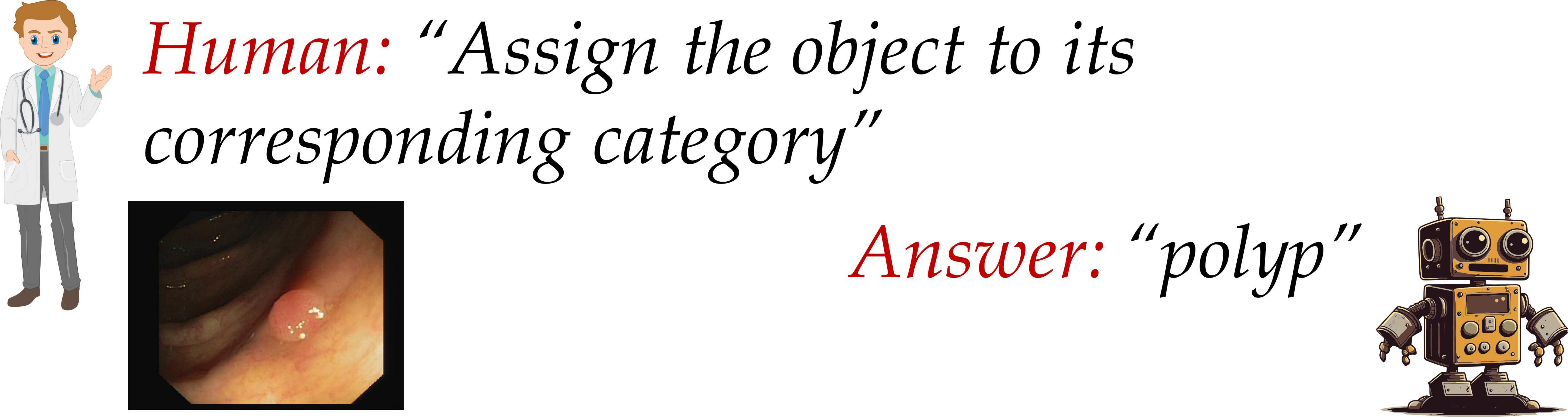}} 
\\
\hline
REG & \makecell[l]{
    1. What category does \{\textit{object coordinates}\} belong to? \\
    2. Can you tell me the category of \{\textit{object coordinates}\}? \\
    3. Could you provide the category for \{\textit{object coordinates}\}? \\
    4. Please specify the category of \{\textit{object coordinates}\}. \\
    5. What is the category for \{coordinates\}?} 
& \makecell[l]{11 sources $\rightarrow$ 
    SUN-database \cite{misawa2021development}, PolypGen \cite{ali2021polypgen}, CVC-ClinicDB \cite{bernal2015wm},\\
    ETIS \cite{silva2014toward}, KUMC \cite{li2021colonoscopy}, Kvasir \cite{pogorelov2017kvasir}, PSNBI2K \cite{psnbi2k}, CVC-ColonDB \cite{bernal2012towards},\\
    EDD2020 \cite{ali2020endoscopy}, Kvasir-Capsule \cite{smedsrud2021kvasir}, Kvasi-Instrument \cite{jha2021kvasir}} 
& \adjustbox{valign=m}{\includegraphics[height=1.15cm]{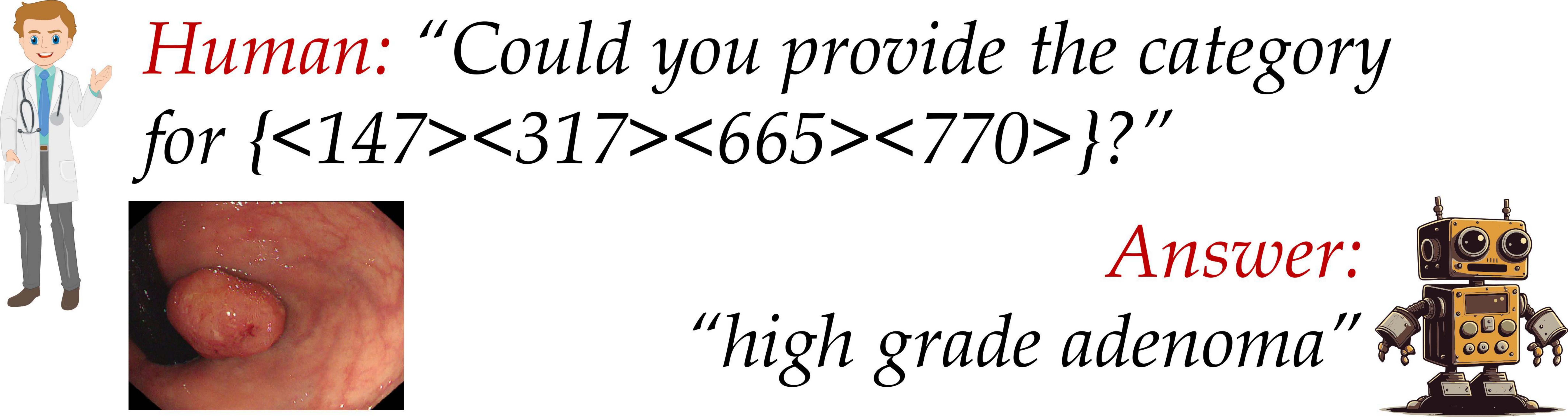}} 
\\
\hline
REC 
& \makecell[l]{
    1. Where is the location of \{\textit{object category}\}? \\
    2. Could you give the position of \{\textit{object category}\}? \\
    3. Where is \{category\} located? \\
    4. Could you specify the location of \{\textit{object category}\}? \\
    5. Please specify the coordinates of \{\textit{object category}\}.} 
& \makecell[l]{11 sources $\rightarrow$ 
    SUN-database \cite{misawa2021development}, PolypGen \cite{ali2021polypgen}, CVC-ClinicDB \cite{bernal2015wm},\\
    ETIS \cite{silva2014toward}, KUMC \cite{li2021colonoscopy}, Kvasir \cite{pogorelov2017kvasir}, PSNBI2K \cite{psnbi2k}, CVC-ColonDB \cite{bernal2012towards},\\
    EDD2020 \cite{ali2020endoscopy}, Kvasir-Capsule \cite{smedsrud2021kvasir}, Kvasi-Instrument \cite{jha2021kvasir}}
& \adjustbox{valign=m}{\includegraphics[height=1.15cm]{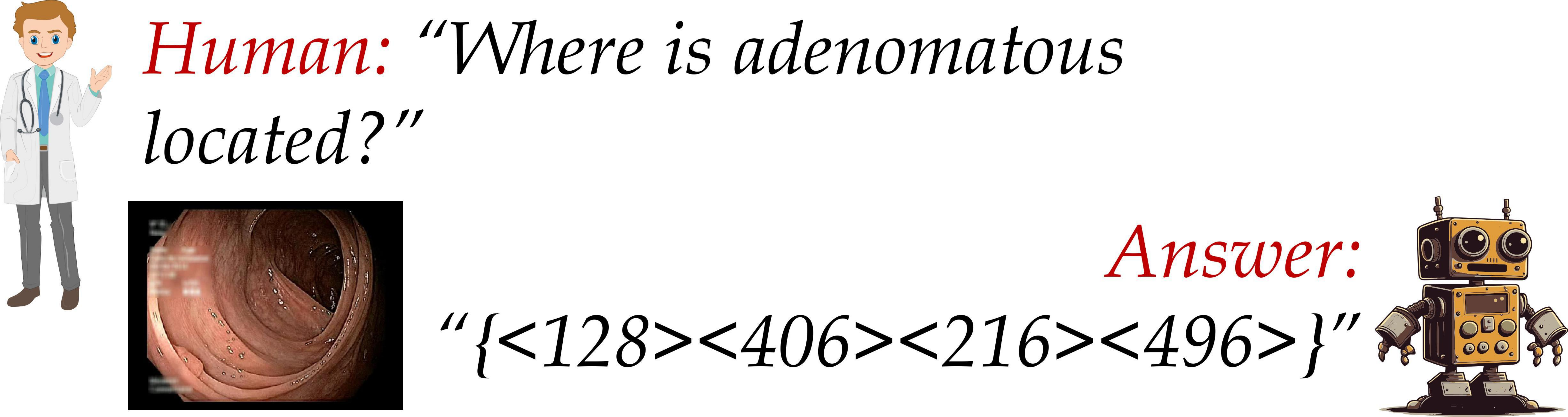}} 
\\
\hline
CAP 
& \makecell[l]{
    1. Describe what you see in the image. \\
    2. Interpret what the image shows. \\
    3. Detail the visual elements in the image. \\
    4. Explain the image's visuals thoroughly. \\
    5. Offer a thorough explanation of the image.}
& \makecell[l]{19 sources $\rightarrow$
    SUN-database \cite{misawa2021development}, PolypGen \cite{ali2021polypgen}, CVC-ClinicDB \cite{bernal2015wm},\\
    ETIS \cite{silva2014toward}, KUMC \cite{li2021colonoscopy}, Kvasir \cite{pogorelov2017kvasir}, PSNBI2K \cite{psnbi2k}, CVC-ColonDB \cite{bernal2012towards},\\
    EDD2020 \cite{ali2020endoscopy}, Kvasir-Capsule \cite{smedsrud2021kvasir}, CP-CHILD \cite{wang2020improved}, 
    BKAI-Small \cite{ngoc2021neounet},\\ 
    PICCOLO \cite{sanchez2020piccolo}, WCE-CCDD\cite{montalbo2022diagnosing}, CPC-Paired \cite{wang2021colorectal}, HyperKvasir \cite{Borgli2020}, \\
    Nerthus \cite{pogorelov2017nerthus}, GastroVision \cite{jha2023gastrovision}, Kvasi-Instrument \cite{jha2021kvasir}}
&\adjustbox{valign=m}{\includegraphics[height=1.15cm]{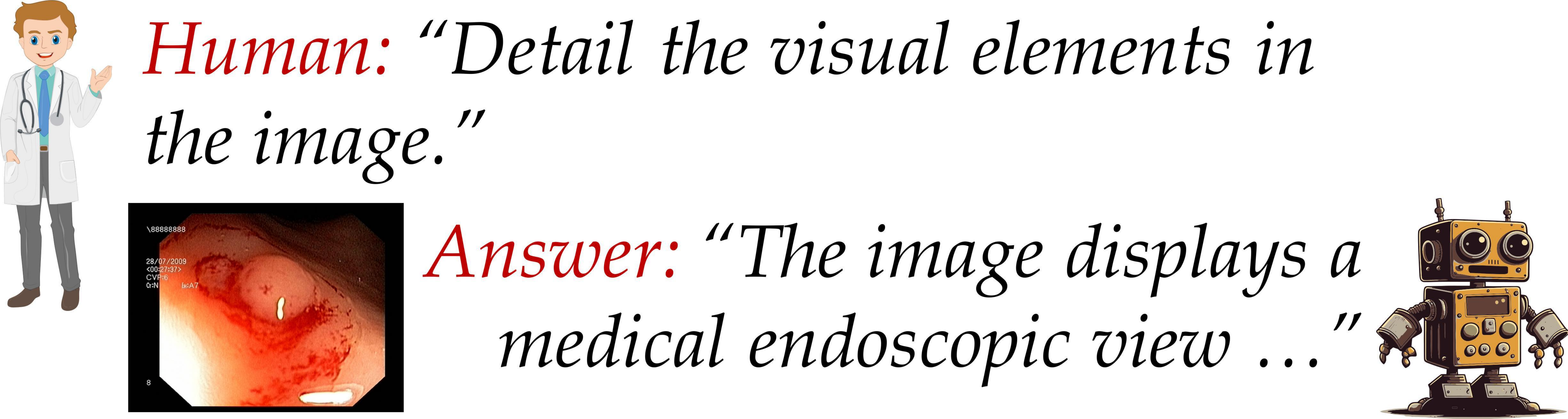}} 
\\
\end{tabular}
\end{table*}

Recently, MLMs have demonstrated significant promise in leveraging language models to process multimodal signals, especially in ``perceiving and interpreting'' visual signals. Instruction tuning \cite{li2024vision} is key to instructing the MLMs to execute domain-specific tasks aligned with user preferences.
This section introduces three initiatives to advance multimodal research: how we create a large-scale instruction tuning dataset, \ourdata~(\secref{sec:colon300k}) and how we train a colonoscopy-specific MLM, {\ourmodel} (\secref{sec:colongpt}). Finally, we contribute the first multimodal benchmark for conversational tasks (\secref{sec:multimodal_benchmark}), conduct diagnostic studies on \ourmodel~(\secref{sec:diagnostic_experiments}), and share lessons from our empirical observations (\secref{sec:empirical_takeaways}).

\subsection{Established instruction tuning dataset: \ourdata}\label{sec:colon300k}

Fig. \hyperlink{fig4_a}{(4-a)} depicts our semi-automated workflow to create instruction tuning data in three steps. We begin by assembling colonoscopy images from public datasets. Following this, we use a category-specific prompt to interact with a multimodal AI chatbot, GPT-4V \cite{openai2023chatgpt4v}, yielding descriptive medical captions for these positive cases within the assembled data. Lastly, we reorganise the instruction tuning pairs derived from the data afore-prepared, enabling the model to perform four different colonoscopy tasks in an interactive, dialogue-based manner.

\noindent$\bullet$~\textbf{Data collection.}
As shown in Fig. \hyperlink{fig4_b}{(4-b)}, \ourdata~contains 303,001 colonoscopy images, including 128,620 positive and 174,381 negative cases collected from 19 different data sources. To ensure data integrity and avoid label leakage, we establish a series of management rules to divide each dataset. For datasets with predefined divisions, such as KUMC \cite{li2021colonoscopy}, PICCOLO \cite{sanchez2020piccolo}, WCE-CCDD \cite{montalbo2022diagnosing}, BKAI-Small \cite{ngoc2021neounet}, CP-CHILD \cite{wang2020improved}, Kvasir-Instrument \cite{jha2021kvasir}, and PS-NBI2K \cite{psnbi2k}, we follow their original division rules. When such predefined rules are not available, we conform to widely recognised benchmarks, such as CVC-ClinicDB \cite{bernal2015wm}, CVC-ColonDB \cite{bernal2012towards}, ETIS-Larib \cite{silva2014toward}, and the polyp category in Kvasir \cite{pogorelov2017kvasir} according to the benchmark by Fan \etal \cite{fan2020pranet}, as well as positive samples in SUN-database \cite{misawa2021development} following the benchmark by Ji \etal \cite{ji2022video}.
For the remaining datasets (HyperKvasir \cite{Borgli2020}, Kvasir-Capsule \cite{smedsrud2021kvasir}, CPC-Paired \cite{wang2021colorectal}, Nerthus \cite{pogorelov2017nerthus}, GastroVision \cite{jha2023gastrovision}, EDD2020 \cite{ali2020endoscopy}, PolypGen \cite{ali2021polypgen}, negative samples in SUN-database \cite{misawa2021development}, remaining categories in Kvasir \cite{pogorelov2017kvasir}), we allocate them proportionally as 60\%/10\%/30\% for training/validation/testing purposes. With the above management rules, our final image division for is roughly 59.73\%/8.77\%/31.61\%. 
As depicted in Fig. \hyperlink{fig4_c}{(4-c)}, all images are classified into a three-level structure, including 4 root/13 parent/62 child categories. In detail, the root level contains three positive categories: the pathological findings of various colonic diseases (110,970 cases); the anatomical structure related to the colon (5,511 cases); and therapeutic interventions related to colonoscopy (12,139 cases), covering the pre-operative, intra-operative, and post-operative stages. 
Targets outside our interest (\eg, stomach, oesophagus, normal Z-line, and gastric polyp not occurred during a colonoscopy) or lack of objects (\eg, normal mucosa, colon background) are classified under the negative root category (174,381 cases). We intentionally keep these negative samples, as they may be valuable for future gastrointestinal research. As shown in Fig. \hyperlink{fig4_d}{(4-d)}, we present a word cloud distribution for all categories names within \ourdata.


\noindent$\bullet$~\textbf{Caption generation.} 
The behavioural study \cite{lupyan2020effects} suggests that language affects both higher-level (recognition) and lower-level (discrimination) processes of visual perception. This encourages us to extend positive cases (\ie, 128,620 images featuring various colonoscopic findings) with descriptive captions. For this purpose, a straightforward way to create captions is to wrap the category name in a basic template, like ``a photo of a [category]'' as used by Radford \etal~\cite{radford2021clip}. However, these simple sentences tend to yield suboptimal multimodal alignment, as they are less informative. As shown in Fig. \hyperlink{fig4_e}{(4-e)}, we introduce a pipeline to generate more descriptive captions. We interact with GPT-4V \cite{openai2023chatgpt4v} using a custom prompt for each colonoscopy image. These prompts act as a prior, conditioned by the image's category. Compared to simple sentences, the generated captions explain three features. First, our generated captions describe the unique patterns of the target, such as ``smooth, round, and nodular appearance'', providing details of the surface of the object. Second, conditioned by category priors, our captions can better differentiate between benign hyperplastic polyps and malignant lesions, describing lesion as ``surface appears intact without evidence of erosion or ulceration''. Third, our captions cover not only the lesion itself but also the surroundings, \eg, ``mucosa appears slightly erythematous, but without significant inflammatory changes'', offering a more holistic and accurate interpretation.

\noindent$\bullet$~\textbf{Organising tuning data.}
In the final step, we convert all positive cases into the single-round conversational format, \ie, ``image+human instruction$\rightarrow$machine response''. As depicted in Fig. \hyperlink{fig4_f}{(4-f)}, we reorganise 450,724 human-machine dialogue pairs from various image/label/caption sources.
Specifically, the classification task (CLS) requires the model to assign a category tag to a colonoscopy image. Using the localisation labels, we introduce two related tasks: referring expression generation (REG), which involves classifying a specified image region, and referring expression comprehension (REC), which involves locating an object with the given category. We also introduce the image captioning (CAP) task that uses GPT-4V-generated captions as machine responses. To enhance dialogue diversity, we set up five question templates per task, in which we randomly select one to form a human-machine dialogue pair, explained in \tabref{tab:coloninst_dialogue}. 

\subsection{Proposed multimodal language model: \ourmodel}\label{sec:colongpt}

\begin{figure}[t!]
\centering
\includegraphics[width=\linewidth]{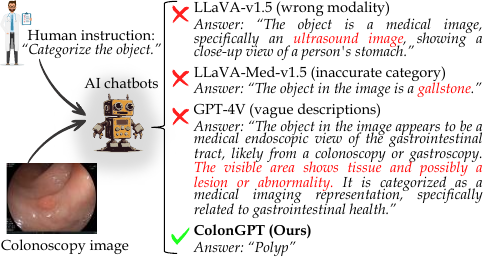}
\vspace{-15pt}
\caption{\textbf{Response comparison for colonoscopy image classification.} We evaluate the zero-shot language responses from three AI chatbots against the response from our multimodal model, \ourmodel.
}
\label{fig:zs_comparison}
\end{figure}

\noindent$\bullet$~\textbf{Motivation.} 
As illustrated in \figref{fig:zs_comparison}, three AI chatbots -- LLaVA-v1.5 \cite{liu2024llavav15}, LLaVA-Med-v1.5 \cite{li2024llavamed}, and GPT-4V \cite{openai2023chatgpt4v} --
are evaluated for their zero-shot language response capabilities. They often produce inaccurate or vague responses and cannot be readily adaptable to specific colonoscopy tasks. This motivates us to develop a colonoscopy-specific MLM for the community. As a result, {\ourmodel} classifies the image as ``polyp'' according to the user instructions, allowing for more precise and personalised applications.

\begin{figure}[t!]
\centering
\includegraphics[width=\linewidth]{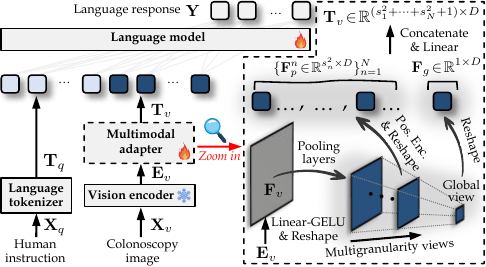}
\vspace{-15pt}
\caption{\textbf{Details of our multimodal language model, \ourmodel.}}
\label{fig:Fig6_ColonGPT}
\end{figure}

\noindent$\bullet$~\textbf{Overview.} 
We strive to verify the efficacy of language models (LMs) in interpreting both visual and textual signals within the field of medical optical imaging. We present a baseline model, called \ourmodel, to execute colonoscopy tasks following human instructions. As shown in the left of \figref{fig:Fig6_ColonGPT}, we follow the framework of MLM \cite{liu2024llavav15}, which typically involves four basic components. 
\textit{(a)} A language tokenizer translates a human instruction $\mathbf{X}_q$ into a sequence of tokens $\mathbf{T}_q\!\in\!\mathbb{R}^{N_q \times D}$, where $N_q$ signifies the length of textual tokens and $D$ represents the embedding dimension. 
\textit{(b)} A visual encoder, typically based on a transformer, condenses a colonoscopy image $\mathbf{X}_v\!\in\!\mathbb{R}^{H \times W \times 3}$, with height $H$ and width $W$, into a flattend visual embedding $\mathbf{E}_v\!\in\!\mathbb{R}^{\frac{HW}{P^2} \times d}$. Here, $P$ denotes the patch size and $d$ refers to the token dimension. 
\textit{(c)} A multimodal adapter transforms the visual embedding $\mathbf{E}_v$ into $N_v$ numeric tokens $\mathbf{T}_v\!\in\!\mathbb{R}^{N_v \times D}$, matching the language dimension $D$ with $\mathbf{T}_q$. 
\textit{(d)} Finally, a language model receives the concatenated visual tokens $\mathbf{T}_v$ and text tokens $\mathbf{T}_q$ as input. Using the chain rule of probability, a sequence $\mathbf{Y}$ of length $L$ is generated in an autoregressive way, formulated as $p(\mathbf{Y}) \!=\! \prod^L_{i=1} p(y_i \mid \mathbf{T}_v,\mathbf{T}_q,\mathbf{Y}_{<i})$,
%
where $\mathbf{Y}_{<i} = [ y_1, y_2, \cdots, y_{i-1} ]$ is the sequence of predicted language tokens indexed before $i$.

\noindent$\bullet$~\textbf{Multigranularity multimodal adapter.} Previous works \cite{chen2023minigptv2,liu2024llavav15,he2024bunny} generally employ a multilayer perceptron architecture as a multimodal adapter, typically consisting of triple linear layers with intervening GELUs. However, handling all visual tokens introduces redundancy because not every token is equally significant, and it also incurs higher computational costs given the quadratic complexity in relation to the number of input tokens. To embrace these challenges, we propose a multimodal adapter that incorporates multigranularity pooling layers between two linear layers. As illustrated in the right of \figref{fig:Fig6_ColonGPT}, we transform the embedding $\mathbf{E}_v\!\in\!\mathbb{R}^{\frac{HW}{P^2} \times d}$ from $d$-dim to $D$-dim using a linear layer followed by a GELU, then reshape it into the spatial format $\mathbf{F}_v\!\in\!\mathbb{R}^{\frac{H}{P} \times \frac{W}{P} \times D}$. To reduce the number of visual tokens while avoiding performance drops, we roll out three modifications for the pooling phase, each validated in \secref{sec:diagnostic_experiments}.
\textit{(a) Multigranularity views.} We add a set of adaptive average pooling operations with $N$ kernel sizes $\{s_1, \dots, s_N\}$ to obtain multigranularity features. In particular, this adaptive operator accommodates input sequences of varying lengths, \ie, the pooled feature for the kernel size $s_n$ is shaped as $\mathbb{R}^{s_n \times s_n \times D}$. \textit{(b) Positional encoding.} Inspired by \cite{islam2019much}, we enhance the spatial information within each pooled feature by applying a 2D convolutional layer with the appropriate zero-padding setting. By default, a zero-pixel boundary is added around the input feature. \textit{(c) Global view.} We also use a global average pooling layer with kernel size $1$ on the feature $\mathbf{F}_v$ to obtain a global view with the shape of $\mathbb{R}^{1 \times 1 \times D}$.
Lastly, we reshape each pooled feature into flattened vectors: $\{\mathbf{F}_p^{n}\!\in\!\mathbb{R}^{s_n^2 \times D}\}_{n=1}^N$ and $\mathbf{F}_{g}\!\in\!\mathbb{R}^{1 \times D}$. We concatenate these vectors and process the resulting vector through the second linear layer to produce the final visual tokens $\mathbf{T}_v\!\in\!\mathbb{R}^{N_v \times D}$, where $N_v=(s_1^2+\cdots+s_N^2+1)$ is the length of visual tokens.

\noindent$\bullet$~\textbf{Implementation.} Our model can be integrated with modern off-the-shelf visual encoders and LMs.
To improve reproducibility for average users, we implement \ourmodel\! in a resource-friendly way. First, we employ SigLIP-SO (0.4B parameters) \cite{zhai2023sigmoid} as the visual encoder, with an input resolution of $H\!=\!W\!=\!384$, a patch size of $P\!=\!14$, and a visual embedding dimension of $d\!=\!1152$. This configuration yields a visual embedding $\mathbf{E}_v$ with a shape $\mathbb{R}^{729 \times 1152}$, where the number of visual tokens is $729=\lfloor\frac{384}{14}\rfloor^2$. In addition, Phi-1.5 (1.3B parameters) \cite{li2023textbooks} serves as the language tokenizer and language model, with a embedding size of $D\!=\!2048$. To reduce computational cost, the size of pooling kernels is set to $\{s_1,s_2\}\!=\!\{14,7\}$, significantly reducing the visual tokens $N_v$ from $729$ to $246$, a reduction of $66.26\%$. This design allows us to complete the training in \revA{seven} hours, facilitating rapid proof-of-concept development.

\begin{table*}[t!]
\centering
\scriptsize
\renewcommand{\arraystretch}{0.8}
\renewcommand{\tabcolsep}{1.5mm}
\caption{\textbf{Multimodal benchmark for three conversational tasks.} ``LoRA'' refers to fine-tuning using low-rank adaptation \cite{hu2022lora}. ``EXT'' indicates the use of pre-trained weights on extra general data. We compare the results on the seen samples from the validation set and the unseen samples from the testing set of \ourdata. Symbol $\uparrow$ means that a higher score reflects better performance. 
}
\label{tab:multimodal_lm_benchmark}
\begin{tabular}{r|c|c|c|cc|cc|cc|cc}
    & Visual encoder & Language model &&& 
    &\multicolumn{2}{c|}{CLS task ($\mathcal{A} \uparrow$)} 
    &\multicolumn{2}{c|}{REG task ($\mathcal{A} \uparrow$)} 
    &\multicolumn{2}{c}{REC task (IoU $\uparrow$)} 
    \\
    \cline{7-12}
    Model & (input shape/URL) & (model size/URL) &No. &LoRA &EXT 
    & \textit{seen} & \textit{unseen} 
    & \textit{seen} & \textit{unseen}  
    & \textit{seen} & \textit{unseen} 
    \\
    \hline
    MiniGPT-v2 \cite{chen2023minigptv2} 
    & EVA-G/14 (448px/\href{https://storage.googleapis.com/sfr-vision-language-research/LAVIS/models/BLIP2/eva_vit_g.pth}{link})
    & LLaMA2 (7B/\href{https://huggingface.co/meta-llama/Llama-2-7b-chat-hf}{link})
    & \makecell{\#A$_1$ \\ \#A$_2$} 
    & \makecell{\checkmark \\ \checkmark} 
    & \makecell{           \\ \checkmark} 
    & \makecell{91.49\% \\ 90.00\%}
    & \makecell{77.93\% \\ 76.82\%} 
    & \makecell{94.69\% \\ 87.65\%}
    & \makecell{72.05\% \\ 70.23\%} 
    & \makecell{23.45\% \\ 27.97\%}
    & \makecell{15.36\% \\ 31.13\%}
    \\
    \hline
    LLaVA-v1 \cite{liu2024llavav1} 
    & CLIP-L/14 (224px/\href{https://huggingface.co/openai/clip-vit-large-patch14}{link})
    & Vicuna-v1.3 (7B/\href{https://huggingface.co/lmsys/vicuna-7b-v1.3}{link})
    & \makecell{\#B$_1$ \\ \#B$_2$} 
    & \makecell{\checkmark \\ \checkmark} 
    & \makecell{           \\ \checkmark} 
    & \makecell{87.86\% \\ 89.61\%}
    & \makecell{72.08\% \\ 42.17\%} 
    & \makecell{84.55\% \\ 86.87\%}
    & \makecell{68.11\% \\46.85\%} 
    & \makecell{20.05\% \\ 21.81\%}
    & \makecell{12.72\% \\ 3.24\%}
    \\ 		
    \hline
    LLaVA-v1.5 \cite{liu2024llavav15} 
    & CLIP-L/14 (336px/\href{https://huggingface.co/openai/clip-vit-large-patch14-336}{link})
    & Vicuna-v1.5 (7B/\href{https://huggingface.co/lmsys/vicuna-7b-v1.5}{link})
    & \makecell{\#C$_1$ \\ \#C$_2$} 
    & \makecell{\checkmark \\ \checkmark} 
    & \makecell{           \\ \checkmark} 
    & \makecell{92.97\% \\ 93.33\%}
    & \makecell{79.10\% \\ 80.89\%} 
    & \makecell{98.58\% \\ 99.32\%}
    & \makecell{70.38\% \\ 72.88\%} 
    & \makecell{55.72\% \\ 61.97\%}
    & \makecell{34.32\% \\ 42.31\%} 
    \\
    \hline
    Bunny-v1.0-3B \cite{he2024bunny} 
    & \makecell{SigLIP-SO (384px/\href{https://huggingface.co/google/siglip-so400m-patch14-384}{link})}
    & Phi2 (2.7B/\href{https://huggingface.co/microsoft/phi-2}{link})
    & \makecell{\#D$_1$ \\ \#D$_2$} 
    & \makecell{\checkmark \\ \checkmark} 
    & \makecell{           \\ \checkmark } 
    & \makecell{91.16\% \\ 92.47\%}
    & \makecell{75.50\% \\ 79.50\%} 
    & \makecell{96.61\% \\ 96.02\%}
    & \makecell{69.45\% \\ 75.08\%} 
    & \makecell{46.24\% \\ 54.00\%}
    & \makecell{31.24\% \\ 41.48\%} 
    \\ 		
    \hline
    MGM-2B \cite{li2024minigemini} 
    &\makecell{CLIP-L/14 (336px/\href{https://huggingface.co/openai/clip-vit-large-patch14-336}{link}) \& \\ ConvNeXt-L (768px/\href{https://huggingface.co/laion/CLIP-convnext_large_d_320.laion2B-s29B-b131K-ft-soup}{link})} 
    & Gemma (2B/\href{https://huggingface.co/google/gemma-2b-it}{link})
    & \makecell{\#E$_1$ \\ \#E$_2$} 
    & 
    & \makecell{ \\ \checkmark} 
    & \makecell{92.97\% \\ 93.24\%}
    & \makecell{78.99\% \\ 78.69\%} 
    & \makecell{98.17\% \\ 98.75\%}
    & \makecell{69.81\% \\ 74.30\%} 
    & \makecell{39.78\% \\ 57.25\%}
    & \makecell{16.00\% \\ 25.23\%} 
    \\
    \hline
    MobileVLM-1.7B \cite{chu2023mobilevlm} 
    & CLIP-L/14 (336px/\href{https://huggingface.co/openai/clip-vit-large-patch14-336}{link})
    & MobileLLaMA (1.4B/\href{https://huggingface.co/mtgv/MobileLLaMA-1.4B-Chat}{link})
    & \makecell{\#F$_1$ \\ \#F$_2$}  
    & \makecell{           \\ \checkmark} 
    & \makecell{\checkmark \\ \checkmark} 
    & \makecell{93.02\% \\ 93.64\%}
    & \makecell{78.75\% \\ 80.44\%} 
    & \makecell{97.78\% \\ 97.87\%} 
    & \makecell{73.14\% \\ 78.03\%} 
    & \makecell{47.30\% \\ 51.36\%} 
    & \makecell{31.46\% \\ 34.80\%} 
    \\
    \hline
    LLaVA-Med-v1.0 \cite{li2024llavamed} 
    & CLIP-L/14 (224px/\href{https://huggingface.co/openai/clip-vit-large-patch14}{link}) 
    & LLaMA1 (7B/\href{https://huggingface.co/docs/transformers/main/model_doc/llama}{link}) 
    & \makecell{\#G$_1$ \\ \#G$_2$}  
    & 
    & \makecell{ \\ \checkmark} 
    & \makecell{93.52\% \\ 93.84\% } 
    & \makecell{78.04\% \\ 77.38\%}  
    & \makecell{97.74\% \\ 97.35\%}  
    & \makecell{75.07\% \\ 75.25\%}  
    & \makecell{41.60\% \\ 39.43\%}  
    & \makecell{24.89\% \\ 20.85\%}  
    \\
    \hline
    LLaVA-Med-v1.5 \cite{li2024llavamed} 
    & CLIP-L/14 (224px/\href{https://huggingface.co/openai/clip-vit-large-patch14}{link}) 
    & Mistral-v0.2 (7B/\href{https://huggingface.co/mistralai/Mistral-7B-Instruct-v0.2}{link})
    & \makecell{\#H$_1$ \\ \#H$_2$} 
    & \makecell{\checkmark \\ \checkmark} 
    & \makecell{ \\ \checkmark} 
    & \makecell{93.62\% \\ 87.22\%} 
    & \makecell{79.24\% \\ 66.51\%} 
    & \makecell{99.30\% \\ 90.40\%} 
    & \makecell{73.05\% \\ 70.00\%} 
    & \makecell{64.69\% \\ 13.39\%} 
    & \makecell{41.97\% \\ 12.95\%} 
    \\
    \hline
    \rowcolor{mygray5}
    \textbf{\ourmodel~(Ours)} 
    & SigLIP-SO (384px/\href{https://huggingface.co/google/siglip-so400m-patch14-384}{link})
    & Phi1.5 (1.3B/\href{https://huggingface.co/microsoft/phi-1_5}{link})
    & -
    & \checkmark 
    & 
    & \textbf{94.06\%} & \textbf{83.24\%}
    & \textbf{99.96\%} & \textbf{80.18\%}
    & \textbf{85.74\%} & \textbf{56.24\%}
    \\
\end{tabular}
\end{table*}

\noindent$\bullet$~\textbf{Training recipe.} We implement our model using the PyTorch library, accelerated by \revA{two NVIDIA H200 GPUs}. The AdamW optimiser is used with an initial learning rate of 2e-3 and a cosine learning rate scheduler. \revA{Our reproducible process consists of two training stages.
\textit{(a) Pre-alignment stage:} First, we train the multigranularity multimodal adapter on $\sim$83K image-caption pairs to pre-align the visual encoder with the language model, enabling it to interpret visual tokens. At this stage, both the visual encoder and the LM remain frozen, while the focus is on the multimodal adapter with a learning rate of 2e-4.
\textit{(b) Supervised fine-tuning stage:} Next, we adapt the model to address colonoscopy-related tasks by fine-tuning with CLS, REG, and REC data, amounting to $\sim$202K image-text pairs. For efficiency, the low-rank adaptation strategy (LoRA \cite{hu2022lora}) is applied to the LM, with a rank of $r\!=\!128$ and a scaling factor of $\alpha\!=\!256$. During this stage, we train the adapter and LM with a learning rate of 2e-3 and 2e-4, respectively. 
Each stage lasts three epochs with a batch size of 16 per GPU and gradient accumulation every two steps. The duration of training takes about 1.4 hours for pre-alignment and 5.3 hours for fine-tuning.}

\subsection{Multimodal benchmark}\label{sec:multimodal_benchmark}

\noindent$\bullet$~\textbf{Model competitors.} 
To establish a widely accepted multimodal benchmark for the community, we select eight popular MLMs as competitors, including six general-purpose and two medically tailored models. As shown in \tabref{tab:multimodal_lm_benchmark}, each competitor has two training setups depending on whether it uses LoRA \cite{hu2022lora} or initialises knowledge \revA{from weights pre-trained on additional general data. We retrain each competitor using all the training and validation dialogues from \ourdata.}

\noindent$\bullet$~\textbf{Evaluation protocols.} 
We quantitatively evaluate three conversational tasks for the multimodal benchmark. For the two classification-based tasks, namely CLS and REG, we adopt the accuracy metric ($\mathcal{A}$) to calculate the ratio of correctly predicted categories to the total number of predictions. For the REC task, we use the intersection over union (IoU) metric to measure the localisation precision. Furthermore, due to the subjective nature of language in the CAP task, we qualitatively analyse the medical accuracy of the responses by verifying the correct identification of the anatomical structures and category names visible in the images, or relevant clinical descriptions.

\noindent$\bullet$~\textbf{Learning ability.}
We begin by conducting an open-book test for each model to quantitatively measure how effectively each model has internalised the visual and linguistic patterns from the training phase. Specifically, we evaluate each model on the samples they have seen during training, \ie, validation dialogues in \ourdata.
The ``\textit{seen}'' columns in \tabref{tab:multimodal_lm_benchmark} show that we achieve the highest scores in the \revA{CLS ($\mathcal{A}\!=\!94.06\%$), REG ($\mathcal{A}\!=\!99.96\%$), and REC ($\text{IoU}\!=\!85.74\%$) tasks}. This suggests that our \ourmodel~has better learning ability, which allows it to correctly classify images and understand reference expressions related to specific visual regions.

\noindent$\bullet$~\textbf{Generalisation ability.}
We further conduct a closed-book test to examine each model's ability to generalise knowledge to unseen conditions, \ie, testing samples of \ourdata. The ``unseen'' columns in \tabref{tab:multimodal_lm_benchmark} consistently reveal our superior performance on unseen samples across all three tasks.
\revA{Recall that our model slightly outperforms LLaVA-v1.5 in the REG task for the seen scenarios, \ie, $99.96\%$ (Ours) \textit{vs.} $99.32\%$ (\#C$_2$). However, this gap is further widened when exposed to unseen scenarios, where we achieve an accuracy of $80.18\%$, even exceeding the 7B-level LLaVA-v1.5 (\#C$_2$) by $7.3\%$.
Moreover, compared to medically tailored models, \ourmodel~shows great potential as a colonoscopy-specific conversational assistant for multimodal tasks, especially in generalising to unseen data. For example, we exceed \#G$_1$ and \#H$_1$ models by $5.2\%$ and $4.0\%$, respectively in the unseen scenarios for the CLS task.}

\begin{figure*}[t!]
\centering
\includegraphics[width=\linewidth]{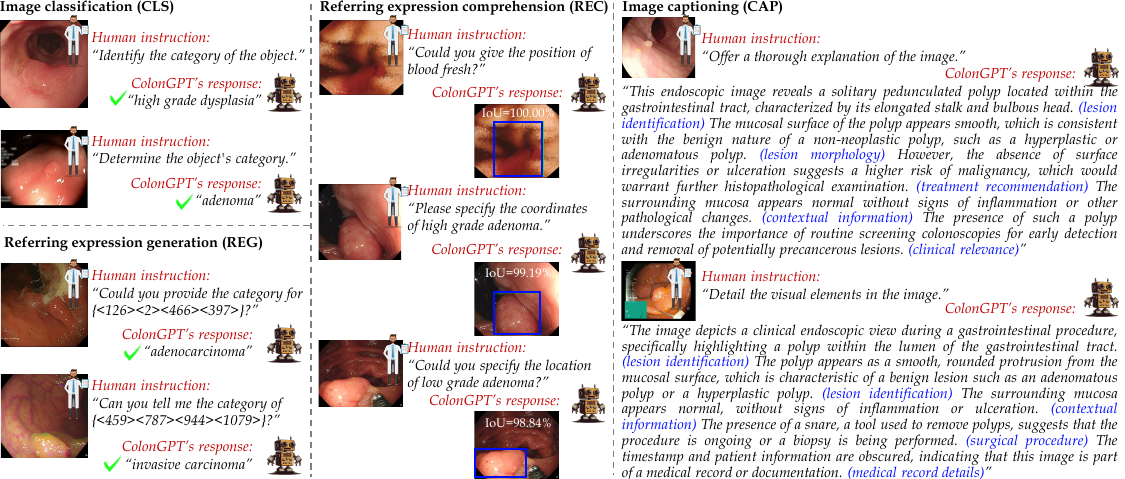}
\vspace{-15pt}
\caption{\textbf{Illustration of \ourmodel's multimodal capabilities.} Our model can execute various multimodal colonoscopy tasks through conversational interactions, including comprehension (CLS, REG), localisation (REC), and captioning (CAP) based.}
\label{fig:qualitative_results}
\end{figure*}

\begin{table*}[t!]
\scriptsize
\renewcommand{\tabcolsep}{0.72mm}
\centering
\caption{\textbf{Diagnostic studies of three core components in \ourmodel.} {``$*$'': interpolate the position embeddings for higher resolution, specifically from 224px to 384px.} Our default configurations are shaded with a gray background.
}
\begin{tabular}{ccc}
\renewcommand{\arraystretch}{1.01}
\renewcommand{\tabcolsep}{0.7mm}
\begin{tabular}{r|r|r|r|r}
\multicolumn{5}{c}{\hypertarget{tab8_a}{\textbf{(a) Different presentations from visual encoder}}} \\
    Visual encoder & input/URL &CLS &REG &REC \\ 
\hline
    ConvNeXtV2-L & 384px/\href{https://huggingface.co/facebook/convnextv2-large-22k-384}{link} & 81.51\% & 76.41\% & 53.09\% \\
    ViT-L & 384px/\href{https://huggingface.co/google/vit-large-patch16-384}{link} &\textbf{83.99\%} &78.79\%	&56.64\% \\
    MAE-L* & 384px/\href{https://huggingface.co/facebook/vit-mae-large}{link} & 83.30\% &79.51\% &53.59\% \\
    MAE-L & 224px/\href{https://huggingface.co/facebook/vit-mae-large}{link} &82.85\% &79.19\% &\textbf{57.57\%} \\ 
    DINOv2-L & 224px/\href{https://huggingface.co/facebook/dinov2-large}{link} &22.48\%	&10.09\% &5.65\% \\
    CLIP-L & 336px/\href{https://huggingface.co/openai/clip-vit-large-patch14-336}{link} &77.05\%	&71.92\%	&51.68\% \\
\hline
    \rowcolor{mygray5}
    SigLIP-SO & 384px/\href{https://huggingface.co/google/siglip-so400m-patch14-384}{link} &83.24\%	&\textbf{80.18\%} &56.24\% \\
\end{tabular}
&
\renewcommand{\arraystretch}{0.9}
\renewcommand{\tabcolsep}{0.7mm}
\begin{tabular}{r|r|r|r|r}
\multicolumn{5}{c}{\hypertarget{tab8_b}{\textbf{(b) Multigranuarity multimodal adapter}}}\\
     &token (ratio) &CLS &REG &REC \\ 
\hline
    MLP baseline & 729 (100.00\%) &78.10\%	&75.27\% &56.29\%	 \\
\hline
    $\{16,8,1\}$ & 321 (44.03\%) &79.84\% &76.13\%	&53.15\% \\
    \rowcolor{mygray5} $\{14,7,1\}$ & 246 (33.74\%) &\textbf{83.24\%} &\textbf{80.18\%} &56.24\% \\
    \rowcolor{white}
    $\{14,7\}$ & 245 (33.61\%) &80.67\%	&78.86\% &\textbf{58.05\%} \\
    $\{12,6,1\}$ & 181 (24.83\%) &79.46\% &75.25\%	&52.22\% \\
    $\{10,5,1\}$ & 126 (17.28\%) &81.42\% &76.73\% &52.04\% \\
    $\{8,4,1\}$ & 81 (11.11\%) &81.55\%	&77.60\% &53.39\% \\
\hline
    \textit{w/o} Pos. Enc. & 246 (33.74\%) & 80.56\%	&76.35\%	&55.77\%    \\
\end{tabular}
&
\renewcommand{\arraystretch}{0.9}
\renewcommand{\tabcolsep}{0.7mm}
\begin{tabular}{r|rr|r|r|r}
\multicolumn{6}{c}{\hypertarget{tab8_c}{\textbf{(c) Fine-tuning strategy}}}\\
    Strategy &$r$ & $\alpha$ &CLS &REG &REC \\ 
\hline
    full-tuning &- &- &70.03\% &63.07\% &13.17\% \\
\hline
    LoRA &8 &16 &80.51\% &77.61\% &51.61\% \\
    LoRA &16 &32 &81.95\% &77.87\% &52.41\% \\
    LoRA &32 &64 &82.63\% &78.76\% &52.11\% \\
    LoRA &64 &128 &82.27\% &79.72\% &52.20\% \\
    \rowcolor{mygray5}
    LoRA &128 &256 &\textbf{83.24\%} &\textbf{80.18\%} &\textbf{56.24\%} \\ 
    \rowcolor{white}
    LoRA &256 &512 &81.75\%	&77.89\% &54.03\% \\
    LoRA &512 &1024 &79.39\% &74.54\% &55.52\% \\
\end{tabular}
\end{tabular}
\label{tab:colongpt_ablation_all}
\end{table*}

\noindent$\bullet$~\textbf{Qualitative analysis.} 
\figref{fig:qualitative_results} illustrates our model's three multimodal abilities across four conversational tasks. \textit{(a) Comprehension ability:} In the CLS task, we identify subtle visual features, distinguishing ``high grade dysplasia'' from ``adenoma'' in visually similar images. In the REG task, we correctly translate complex visual features from the given coordinates into precise medical terminology. \textit{(b) Localisation ability:} This entails \ourmodel~understanding language query and localising visual target within a complex colon environment. The outputs of the REC task showcase \ourmodel's precision in localising specified expressions using bounding boxes. 
\textit{(c) Captioning ability:} This requires the synthesis of visual information into coherent, clinically relevant text. Our model provides descriptions of a pedunculated polyp, detailing its morphology, contextual characteristics, and potential clinical relevance. {Additionally, \ourmodel~can describe the treatment procedure when an instrument is present, \revA{\eg, ``The presence of a snare, a tool used to remove polyp''}
}

\subsection{Diagnostic experiments}\label{sec:diagnostic_experiments}

\noindent$\bullet$~\textbf{Visual encoder.} 
Our diagnostic experiment begins with an inquiry -- \textit{What types of visual representations are appropriate for multimodal colonoscopy data?} We prepare four sets of representations from various large-scale pre-training strategies: supervised learning (ConvNeXtV2 \cite{woo2023convnextv2}, ViT \cite{dosovitskiy2021vit}), reconstructive learning (MAE \cite{he2022masked}), and contrastive learning using vision-only data (DINOv2 \cite{oquab2023dinov2}) or VL data (CLIP \cite{radford2021learning}, SigLIP \cite{zhai2023sigmoid}).
As shown in Tab. \hyperlink{tab8_a}{(VIII-a)}, all encoders use pre-trained weights from Huggingface. To ensure consistency, we manually interpolate the smaller position embedding for MAE and DINOv2 from 224px to 384px (marked with $*$), while leaving the default input of the remaining models unchanged. 
Our observation reveals that contrastive learning encoders using VL data outperform other strategies. This suggests that visual representations pre-aligned with weak texts during their pre-training facilitate us to transform visual embeddings into the language space.
Regarding the other unimodal encoders, both supervised learning methods (ConvNeXtV2, ViT) and the reconstructive approach (MAE) give satisfactory feedback. However, the vision-only contrastive learning model (DINOv2) struggles\footnote{\revA{Notably, multiple efforts to optimise DINOv2 with various hyperparameters, including input size, model size, and learning rate, did not produce acceptable results for multimodal colonoscopy tasks.}}, \revA{suggesting that its visual representations may be difficult to align with the language space.}

\noindent$\bullet$~\textbf{Multigranularity multimodal adapter.} 
It serves as a key component in linking vision and language modalities, reducing visual tokens to mitigate computational overhead. As detailed in Tab. \hyperlink{tab8_b}{(VIII-b)}, we analyse its effectiveness from three perspectives.
\textit{(a) How to configure the pooling kernels?} As a reference, we initialise a baseline variant of {\ourmodel} with a multimodal adapter used in \cite{liu2024llavav15,chen2023minigptv2}, which employs a pure MLP architecture to process all input tokens equally. We then gradually decrease the size of the pooling kernels across five variants: $\{16,8,1\}$, $\{14,7,1\}$, $\{12,6,1\}$, $\{10,5,1\}$, and $\{8,4,1\}$. {Concerning the performance-cost trade-offs, our setup $\{14,7,1\}$ is optimal. It decreases the visual tokens from 100\% (729 tokens) to 33.74\% (246 tokens) while maintaining impressive results across three conversational tasks. To illustrate, we observe \revA{performance gains of 5.14\% and 4.91\% in the CLS and REG tasks}, respectively.}
\textit{(b) Is global context necessary?} We remove the global view from our default setup $\{14,7,1\}$ for the multigranularity adapter, producing a controlled variant with setup $\{14,7\}$. The performance then declines, indicating the necessity of capturing the global context within visual embeddings for improved outcomes.
\textit{(c) Is positional encoding important?} As shown in the last row of Tab. \hyperlink{tab8_b}{(VIII-b)}, our model without positional encoding shows a significant performance drop in the \revA{REG task, from 80.18\% to 76.35\%}. This suggests that the relative position information for the visual sequence is crucial for the localisation task.

\noindent$\bullet$~\textbf{Fine-tuning strategy.} 
Lastly, we investigate \textit{how to effectively tune our model on multimodal colonoscopy data}. As shown in Tab. \hyperlink{tab8_c}{(VIII-c)}, we initiate a set of variants to tune the language model, Phi1.5. It includes seven variants with different LoRA setups and a full-tuning variant as a reference point. The best performance was observed in the LoRA variant with configuration $r/\alpha=128/256$.
\revA{Moreover, the full-tuning variant yields unsatisfactory performance on three tasks. This suggests that an increase in the tunable parameters does not produce gains under the conditions of limited-scale data.}

\subsection{Empirical takeaways}\label{sec:empirical_takeaways}

This study represents a preliminary exploration of multimodal instruction tuning techniques in colonoscopy. We unify the multimodal and multitask paradigms in a causal language model, which features two insights: interpreting visual content within the linguistic space and tackling various visual tasks under a next-token prediction framework. 
We finally derive lessons from experiments that may guide future advances in multimodal research.

\noindent$\bullet$~\textbf{Embracing data scarcity.} 
In general, MLMs \cite{liu2024llavav1,liu2024llavav15} opt for a two-stage strategy trained on massive data, \eg, $\sim$558K samples for multimodal alignment, followed by $\sim$665K instruction tuning samples to ensure human compliance. Alternatively, we adopt a single-stage strategy to directly fine-tune {\ourmodel} on comparatively smaller training data with $\sim$285K instructions. This strategy appears to be effective in colonoscopy, a data-limited scenario. We suggest two feasible ways to compensate for this data-centric issue. 
{\textit{(a)} Scaling up data size is a straightforward way to improve the domain-specific representation ability. A cost-efficient way is to consider synthesised data once the public data sources are used up \cite{villalobos2024position}. \textit{(b)} Diversifying the human-machine dialogue can efficiently train an AI specialist for colonoscopy applications. This involves expanding question-answer pairs with advanced AI chatbots and organising more executable tasks, such as converting masks into polygons for segmentation \cite{xiao2024florence2} or modelling multiframe correlations for video analysis \cite{jiang2024mantis}.}

\noindent$\bullet$~\textbf{Efficiency drives progress.}
As discussed above, we take less
data to obtain greater performance than other model rivals. This success benefits from the way we build \ourmodel.
\textit{(a)} Colonoscopy data inherently contains redundant information, such as the fact that most mucosal surfaces are similar, as well as camouflaged patterns between benign lesions and their surroundings, as discussed in \secref{sec:domain_unique_challenges}. To reduce redundancy, we propose a multigranularity multimodal adapter that selectively samples tokens without compromising performance. For future improvement, we can draw on the wisdom of previous token reduction techniques \cite{haurum2023tokens}.
\textit{(b)} As shown in \tabref{tab:multimodal_lm_benchmark}, the Phi1.5 model \cite{li2023textbooks}, although lightweight, shows surprising efficiency,
even outperforming other 7B-level competitors. This indicates that larger models appear to require more
colonoscopy data. Thus, future efforts should prioritise enhancing \revA{parameter efficiency}, especially for the medical field, rather than racing with massive computational resources. A promising idea to streamline the MLM framework using an encoder-free solution \cite{diao2024unveiling} to interpret visual pixels.

\noindent$\bullet$~\textbf{Improving spatial perception.} We observe that the ability to accurately locate targets given descriptions remains limited. This is evident in the REC results shown in \tabref{tab:multimodal_lm_benchmark}, where IoU scores fall below 50\% in most models when tested on unseen samples. To break through this performance bottleneck, we suggest two potential routes.
\textit{(a)} In constructing \ourmodel, we leverage a pre-trained visual encoder and a language model from the general domain. This approach presents challenges of
the gaps between general and medical optical data, as well as the gap between vision and language modalities. As recommended in \cite{chen2024internvl}, pre-training and pre-aligning the multimodal space before instruction tuning would be a promising approach to alleviate these issues.
\textit{(b)} The next-token prediction framework of causal language models may struggle with arithmetic tasks due to the snowballing error resulting from the chain rule \cite{bachmann2024pitfalls}. For example, LMs are not responsible for accurately predicting coordinates in the REC task. We encourage that the vision and language parts of the next-generation framework can handle their respective roles, such as a parallel framework \cite{huang2024segment} that predicts segmentation masks and generates language captions, simultaneously.

\section{Conclusion}\label{sec:conclusion}

We investigate the frontiers in intelligent colonoscopy techniques and their broader implications in the multimodal field. Our structure follows two primary threads.
First, we survey the landscape of four colonoscopic scene perception tasks and sort out the key challenges and understudied areas. Second, our survey reveals that multimodal research in colonoscopy is underexplored. To this end, we contribute three initiatives to the community: a large-scale multimodal instruction tuning dataset \ourdata, a colonoscopy-specific multimodal language model \ourmodel, and a multimodal benchmark.

\section*{Acknowledgments}
This work was supported by the ANU-Optus BRCoE (scholarship awarded to Ge-Peng Ji) and the NSFC (No. 62476143). We express our sincere gratitude to Yu-Cheng Chou (JHU) and Stephen Gould (ANU) for their interesting discussions.

\bibliographystyle{IEEEtran}
\bibliography{bibliography}

 





\end{document}